\documentclass[preprint,12pt]{aastex}

\shorttitle{Optical/NIR counterpart of IGR~J16318$-$4848}
\shortauthors{Filliatre \& Chaty}

\begin{document}

\title{The optical/NIR counterpart of the INTEGRAL obscured source IGR~J16318$-$4848~: a sgB[e] in a HMXB ?\footnote{Based on observations collected at the European Southern Observatory, Chile (observing proposal ESO N¡ 70.D-0340).}}
\author{P. Filliatre}
\affil{Service d'Astrophysique, CEA/DSM/DAPNIA/SAp}
\affil{CE-Saclay,Orme des Merisiers, B\^{a}t. 709, F-91191 Gif-sur-Yvette Cedex, France}
\affil{F\'{e}d\'{e}ration Astrophysique et Cosmologie, Paris, France}
\email{filliatr@cea.fr}

\and

\author{S. Chaty}
\affil{Universit\'{e} Paris 7, 2 place Jussieu, 75005 Paris}
\affil{Service d'Astrophysique, CEA/DSM/DAPNIA/SAp}
\affil{CE-Saclay,Orme des Merisiers, B\^{a}t. 709, F-91191 Gif-sur-Yvette Cedex, France}
\email{chaty@cea.fr}

\begin{abstract}

The X-ray source IGR~J16318$-$4848 was the first source discovered by INTEGRAL on 2003, January 29. The high energy spectrum exhibits such a high column density that the source is undetectable in X-rays below 2 keV. On 2003, February, 23--25, we triggered our Target of Opportunity (ToO) program using the EMMI and SOFI instruments on the New Technology Telescope of the European Southern Observatory (La Silla) to obtain optical and near-infrared (NIR) observations. We confirm the already proposed NIR counterpart and for the first time extended detection into optical. We report here photometric measurements in the $R$, $I$, $J$, bands, upper flux limits in the $Bb$ and $V$ bands, lower flux limits in the $H$ and $K_{s}$ bands. We also obtain NIR spectroscopy between $0.95$ and $2.52\,\micron$, revealing a large number of emission lines, including forbidden iron lines and P-Cygni profiles, and showing a strong similarity with CI Cam, another strongly absorbed source. Together with the Spectral Energy Distribution (SED), these data point to a high luminosity, high temperature source, with an intrinsic optical-NIR absorption greater than the interstellar absorption, but two orders of magnitude below the X-ray absorption. We propose the following picture to match the data~: the source is a High Mass X-ray binary (HMXB) at a distance between 0.9 and 6.2 kpc, the optical/NIR counterpart corresponds to the mass donor, which is an early-type star, maybe a sgB[e] star, surrounded by a dense and absorbing circumstellar material. This would make the second HMXB with a sgB[e] star as the mass donor after CI Cam. Such sources may represent a different evolutionary state of X-ray binaries previously undetected with the lower energy space telescopes~; if it is so, a new class of strongly absorbed X-ray binaries is being unveiled by INTEGRAL.

\end{abstract}


\section{INTRODUCTION}

IGR~J16318$-$4848 was the first new source discovered by the INTEGRAL IBIS/ISGRI imager \citep{ubertini,lebrun}. The source was first detected on  2003 January 29 \citep{courvoisier} in the energy band 15--40 keV with a mean 20--50 keV flux of $6\times 10^{11}\,\rm erg\,cm^{-2}\,s^{-1}$. The source was thereafter regularly observed for two months. No X-ray counterpart could be found in the ROSAT All Sky Survey \citep{voges}. The source was observed by XMM-Newton on 2003 February 10, which detected a single X-ray source within the INTEGRAL error box using the EPIC PN and MOS cameras \citep{struder,turner}, giving the most precise position to date~: $\alpha=16^{\rm h} 31^{\rm m} 48\fs 6$, $\delta=-48\arcdeg 49\arcmin 00\arcsec$ with a $4\arcsec$ error box (a circle with a radius of $4\arcsec$)\citep{schartel}. X-ray spectroscopy revealed a very high absorption column density : $N_{\rm H}> 10^{24}\,\rm cm^{-2}$ \citep{matt,walter}, which renders the source invisible below 2 keV. This amount of absorption is unusual in Galactic sources. This could explain the non detection by ROSAT, although the source was discovered at a similar flux level in archival ASCA observations in 1994 \citep{murakami,rev1} on both GIS and SIS instruments (between 0.4 and 10 keV). Relatively bright and highly absorbed sources like IGR~J16318$-$4848 could have escaped detection in past X-ray surveys and could still contribute significantly to the Galactic hard X-ray background in the 10--200 keV band.\\
The high column density prompted a counterpart research in the near-infrared (NIR)~: in the EPIC error box, a possible counterpart was proposed by \citet{foschini} using the Two Micron All Sky Survey (2MASS) with the following magnitudes~: $J=10.162\pm0.018$, $H=8.328\pm0.037$, $K_{s}=7.187\pm0.015$ \citep{walter}. On the other hand, no radio emission at the position of the source could be detected. These broad characteristics suggest that it is an X-ray binary, the nature of the compact object (neutron star or black hole) being subject to debate.\\
In the course of a Target of Opportunity (ToO) program at the European Southern Observatory (ESO) dedicated to look for counterparts of high energy sources newly discovered by satellites including INTEGRAL (PI S. Chaty), we carried out photometric observations in the optical and NIR, and  spectroscopic observations in the NIR. The goals were to search for likely counterparts within the EPIC error box, to obtain informations about the environment and the nature of the source, especially about the mass donor.\\
The main results of this paper have been already presented in a very condensed way in \citet{chaty}. In section \ref{secobservations}, we describe briefly our observations and their reduction~; in section \ref{secidentification}, we report on astrometry and photometry of all possible counterparts of the source, and give the most likely candidate~; for the latter, we give in section \ref{secsed} the spectral energy distribution, putting our data together with the survey archives and published data, and derive absorption and temperature estimates~; in section \ref{speclines}, we study the spectral lines~; in section \ref{secdiscuss}, we discuss the distance to the source, its nature and the nature of its components~; then, in section \ref{secconc}, we conclude.

\section{THE OBSERVATIONS}
\label{secobservations}

The observations were carried out using the New Technology Telescope (NTT) of the ESO at La Silla (Chile), on 2003 February 23--25. We used the EMMI (optical) and SOFI (NIR) instruments. For astrometry and photometry, we had a total of 8 broad bands~: $Bb$, $V$, $R$, $I$, $Z$, $J$, $H$, $K_{s}$. For  spectroscopy, we used 2 bands~: GBF (Grism Blue Filter, range $0.95-1.64\,\,\micron$, resolution 930) and GRF (Grism Red Filter, range $1.53-2.52\,\,\micron$, resolution 980).\\
All the data were reduced using the {\it IRAF} suite\footnote{{\it IRAF} is distributed by the National Optical Astronomy Observatories,
    which are operated by the Association of Universities for Research
    in Astronomy, Inc., under cooperative agreement with the National
    Science Foundation.}.

	\subsection{Optical Photometric Data}

	The optical data have been taken on 2003 February 24 with EMMI between UT 7h30 and 8h30. The seeing was typically $0\farcs 9$, and the airmass on the source was between 1.225 and 1.277. We acquired one 180 second exposure for each band ($Bb$, $V$, $R$, $I$, $Z$) and an extra 60 second exposure for $Bb$. The data analysis involved subtraction of the bias and division by the flat field. For calibration, we obtained an exposure of 30, 15, 10, 10, 10 seconds, respectively of the five stars of PG1633+099 of Landolt catalog \citep{landolt}. The EMMI detector is a mosaic of two arrays, each with two readouts, resulting in a $4152\times 4110$ array with a pixel-scale of $0\farcs 166$ per pixel. This makes four parts with slightly different characteristics. The source fell into the second readout of the first array, as did three of the standard stars, therefore we used these three stars to compute the color equation for $BbVRI$. These stars were PG1633+099 B, C and D. As the airmass was identical for these three stars, it was then not possible to determine the extinction coefficient with them, and we adopted the values reported in the EMMI manual \citep{emmiman}. We found the zero-point and the color coefficient by a linear regression. The results are given in Table~\ref{table_system_photom}. The zero-point magnitude is in good agreement with data given in \citet{emmiman} (converted for 1 ADU per second to be directly comparable with our results) : $24.98\pm 0.03$, $25.69\pm 0.01$, $25.98\pm 0.02$ and $25.28\pm 0.03$ for $BbVRI$, respectively.

	\subsection{NIR Photometric Data}

	The NIR photometric data have been taken with SOFI on the NTT between UT 7 and 10~h on 2003 February 23. The SOFI detector is a $1024\times 1024$ CCD with a pixel-scale of $0\farcs 288$ per pixel. The seeing was typically $1\arcsec$, and the airmass on the source was between 1.088 and 1.388. We obtained 9 exposures of 10 seconds and 11 exposures of 5 seconds in the $J$ band. We obtained 5 exposures of 10 seconds and 9 exposures of 2 seconds in the $H$ band. We acquired 245 exposures of 2 seconds in the $K_{s}$ band. In order to correct for the bright NIR sky, following the standard procedure, the pointings of all these frames were dithered. This allowed us to obtain a sky template by median filtering the frames. This template was then subtracted from the flat-fielded frames. The alignment of the frames was performed using $\sim 2000$ common stars. Within each band, frames with different exposure times were averaged separately. For calibration, five exposures for each band of standard stars sj9157 (integration time 2 seconds) and sj9170 (1.2 seconds) of \citet{persson} were done. The color term was taken from \citet{sofiman}, because we had three bands and only two standard stars. We found the zero-point and the extinction coefficient by a linear regression. The results are reported in Table~\ref{table_system_photom}. The zero-point magnitude is in good agreement with data given in \citet{sofiman}~: 23.2, 23.1 and $22.5\pm 0.01$ for $J$, $H$ and $K_{s}$, respectively.

\placetable{table_system_photom}


\begin{deluxetable}{cccc}
\tablecaption{Terms of color equation\label{table_system_photom}. The extinction coefficients for the optical bands are taken from \citet{emmiman}, the color terms for the NIR are taken from \citet{sofiman}, the other coefficients are measured values.}
\tablehead{
\colhead{Band} & \colhead{Zero-point mag.} & \colhead{color term} & \colhead{Extinction coeff.}}
\startdata
$Bb$&$25.03\pm 0.03$&$-0.03\pm 0.01(B-V)$&0.214\tablenotemark{a}\\
$V$&$25.59\pm 0.02$&$ 0.08\pm 0.01(V-R)$&0.125\tablenotemark{a}\\
$R$&$25.90\pm 0.03$&$-0.005\pm 0.2(R-I)$&0.091\tablenotemark{a}\\
$I$&$25.42\pm 0.04$&$ 0.04\pm 0.29(R-I)$&0.051\tablenotemark{a}\\ 
\hline
$J$&$23.34\pm 0.06$&$-0.007(J-K)$\tablenotemark{b}&$0.40\pm 0.05$\\
$H$&$23.17\pm 0.06$&$-0.022(J-K)$\tablenotemark{b}&$0.31\pm 0.05$\\
$K_{s}$&$22.39\pm 0.06$&$ 0.023(J-K)$\tablenotemark{b}&$0.15\pm 0.05$\\
\enddata
\tablenotetext{a}{Not measured, taken from \citet{emmiman}}
\tablenotetext{b}{Not measured, taken from \citet{sofiman}}
\end{deluxetable}


	\subsection{NIR Spectroscopy}

The NIR spectroscopic data have been taken with SOFI low resolution grism on the NTT between UT 8 and 9 h on 2003, February, 25 with the GBF and GRF filters. The seeing was typically $0\farcs 9$, and the airmass on the object varied between 1.160 and 1.260. For each GBF and GRF bands, we obtained 24 exposures of 6 seconds. After subtraction of the bias, the exposures were divided by the response of the flat field along the dispersion line. The resulting individual rough spectra show a dispersion of around 20~\%, which seems to be random, and not associated to a short time variation of the source. Therefore, the two averaged spectra have a relative precision of 4~\% in flux. Given the seeing and the slit width of $1\arcsec$, we expect to collect around 75\% of the flux ; however, as we will scale the spectra to match the broad band magnitudes, absolute calibration is not an issue. Wavelength calibration was done by observing a Xenon arc. We obtained 16 exposures of 5 seconds for standard star hip80456 ($m_{V}=7.74$, spectral type F5V), 8 exposures of 10 seconds for standard star hip83612 ($m_{V}=8.38$, spectral type G1V). With these stars we made a telluric correction, followed by a flux calibration using a blackbody spectrum extrapolated from the magnitude and spectral type of the standard star. Some absorption lines especially Br$\gamma$, were seen on the standard stars : we removed the lines and replaced them by a linear interpolation, as the continuum was smooth enough.
As for the object the two bands did not match in continuum flux in the overlap region, the continuum was extracted and scaled to match the $J$ and $K_{s}$ broadband values, giving a good agreement with the $H$ broadband value (see subsection~\ref{bandmag}). 
In two regions, between 1.35 and $1.45\,\,\micron$, and between 1.8 and $2.0\,\,\micron$, there is a strong absorption and the flux calibration did not work successfully. The spectrum is also very noisy above $2.35\,\,\micron$.

\section{IDENTIFICATION OF THE COUNTERPART}
\label{secidentification}

\subsection{Astrometry of the candidates}

\begin{figure}
\epsscale{0.3}
\plotone{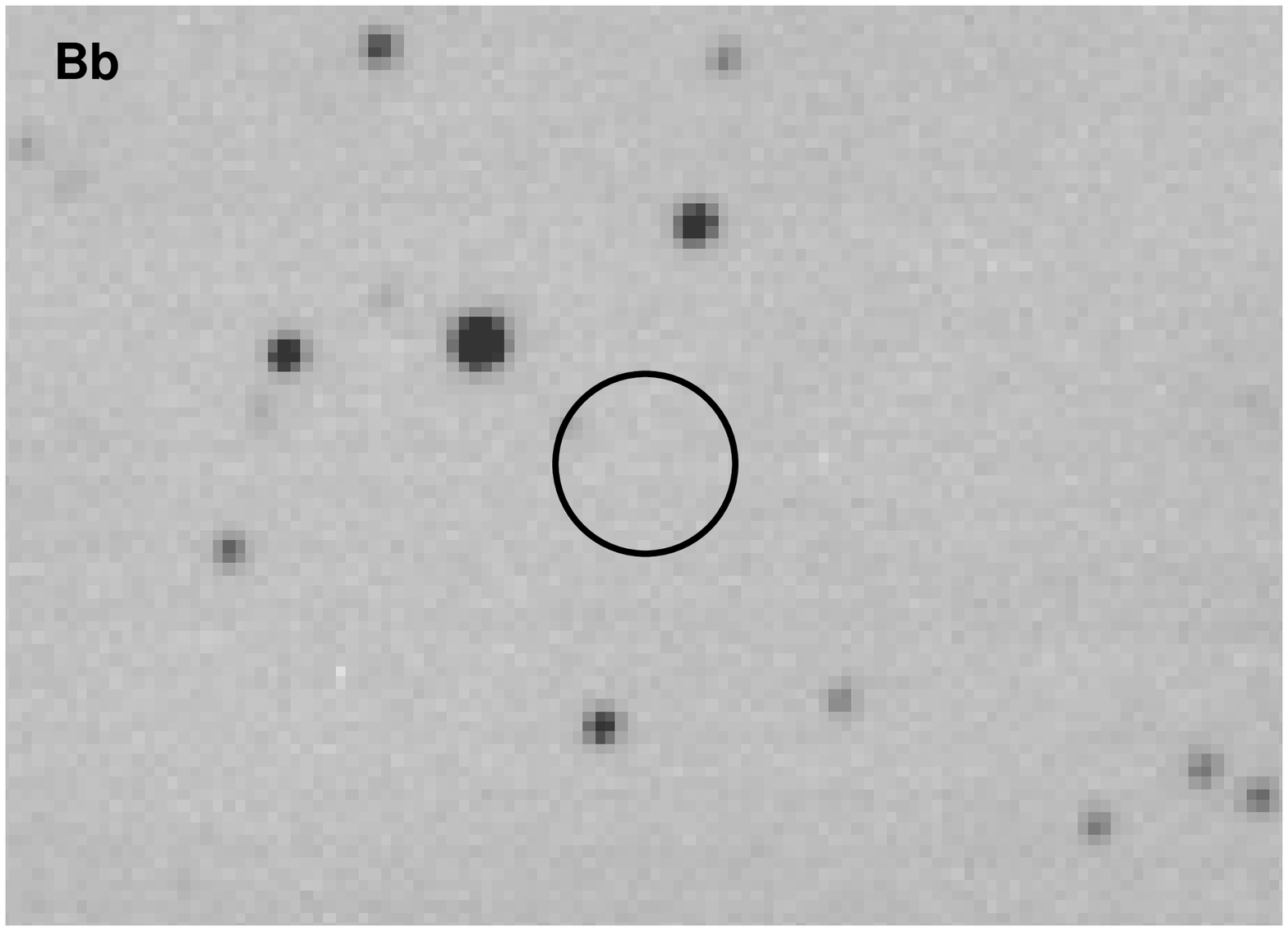}
\plotone{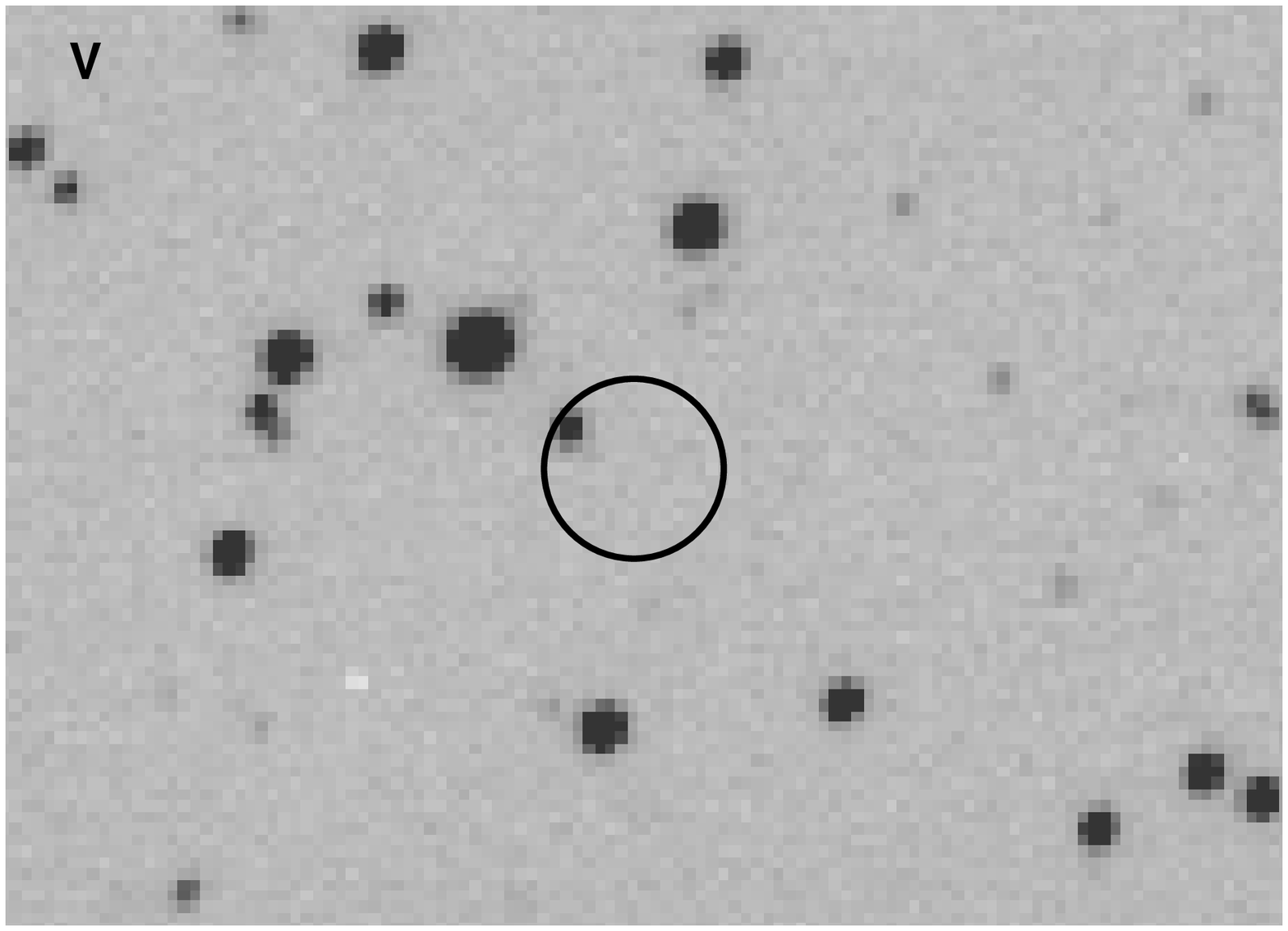}
\plotone{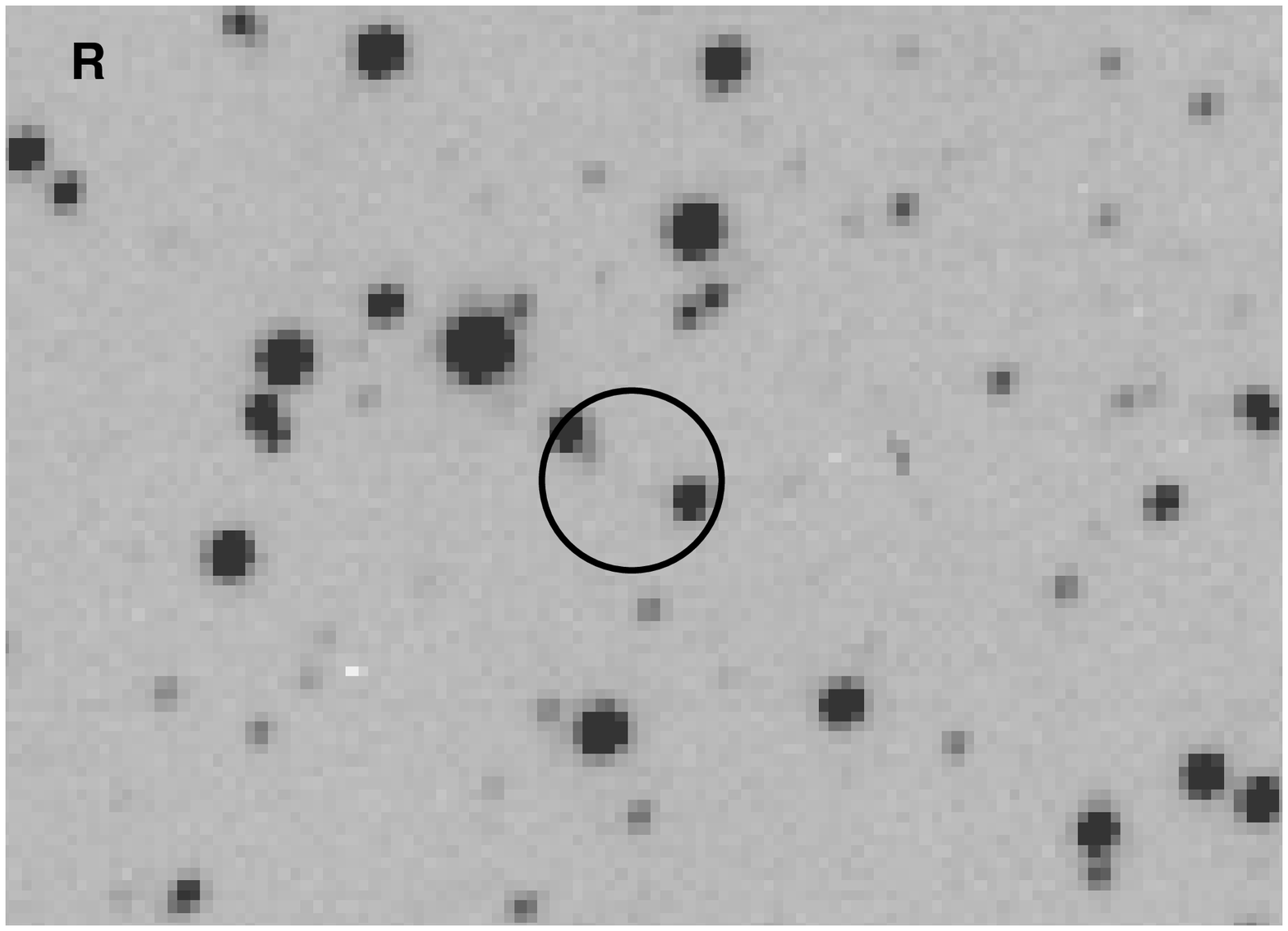}
\plotone{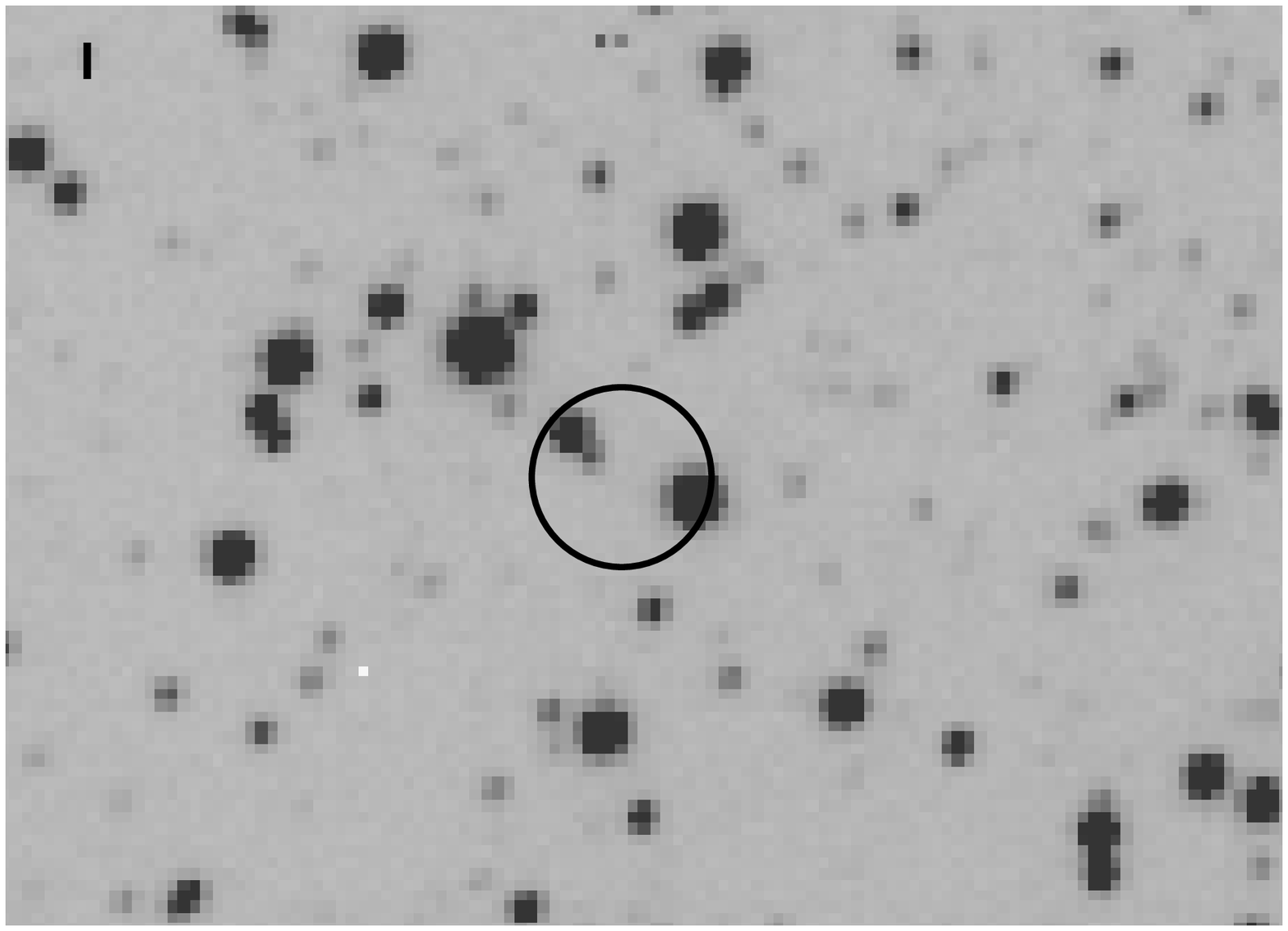}
\plotone{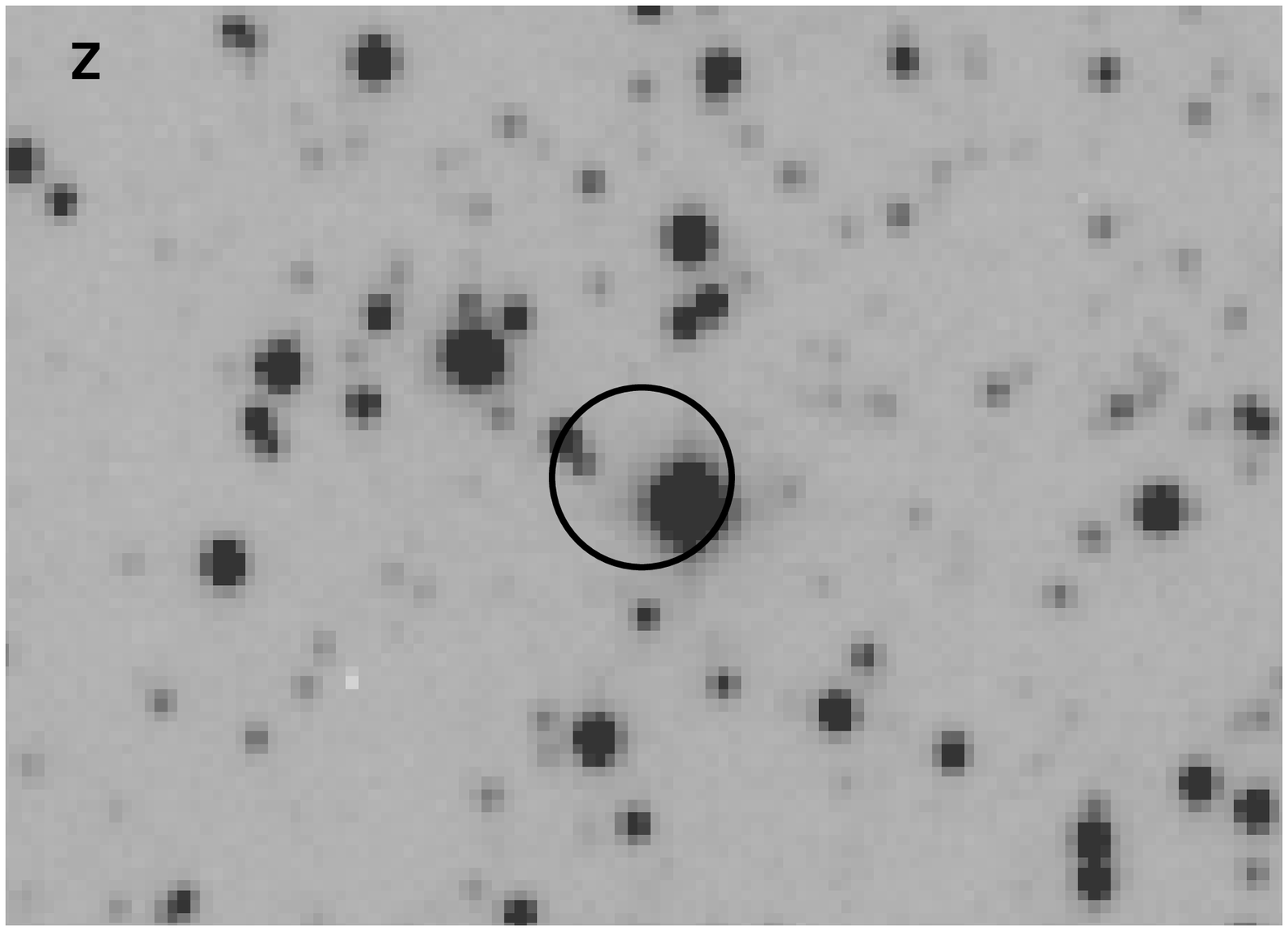}
\caption{The error box of $4\arcsec$ is shown on the $BbVRIZ$ frames of EMMI. North is up, east is left. The scale is given by the error box. Candidate 1 is at the south-west border of the circle, and candidates 2 and 3 are at the north-east border (see Figure~\ref{pos_star_sofi} for identification).}
\label{pos_star_emmi}
\end{figure}

\begin{figure}
\plotone{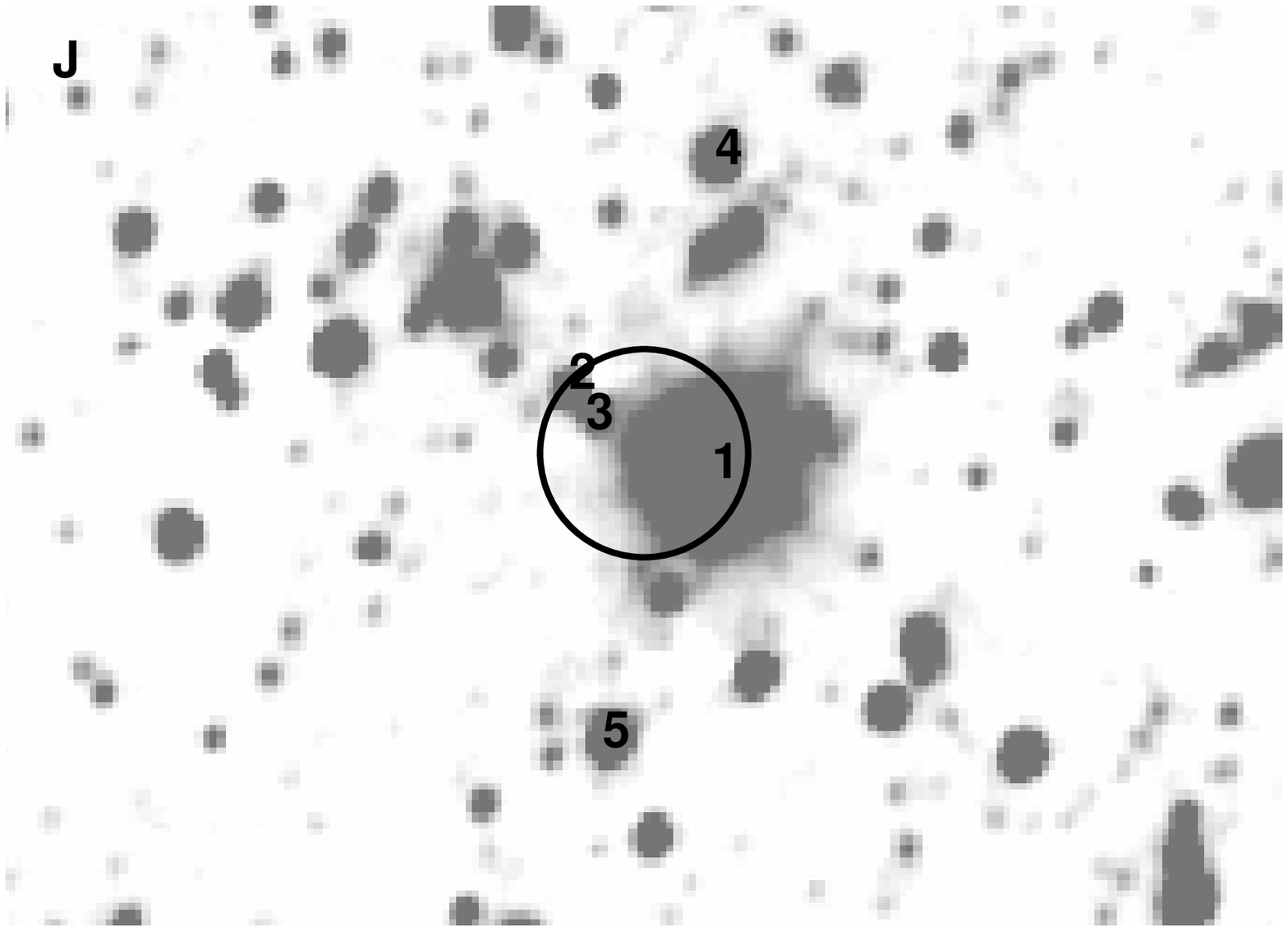}
\plotone{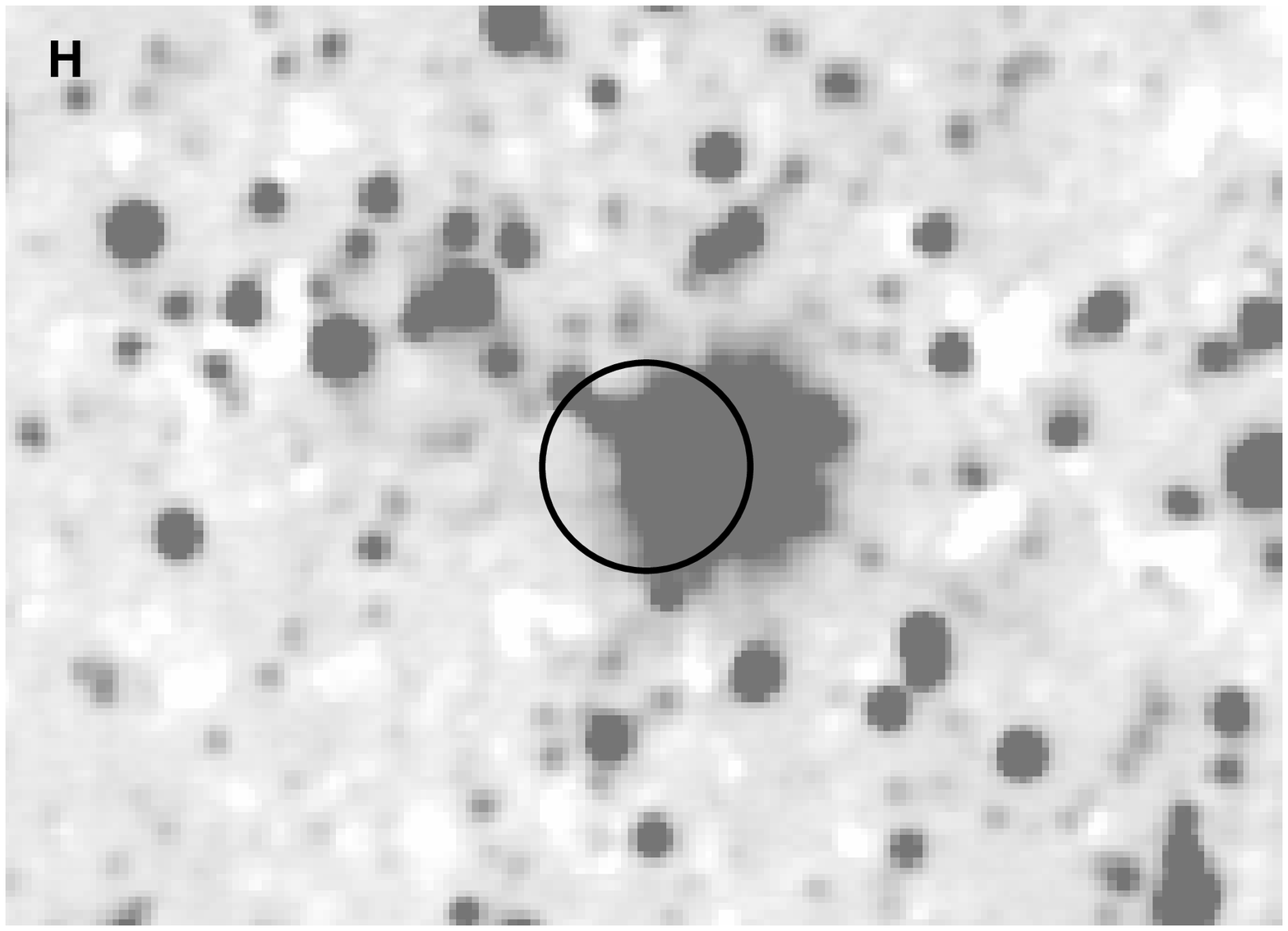}
\plotone{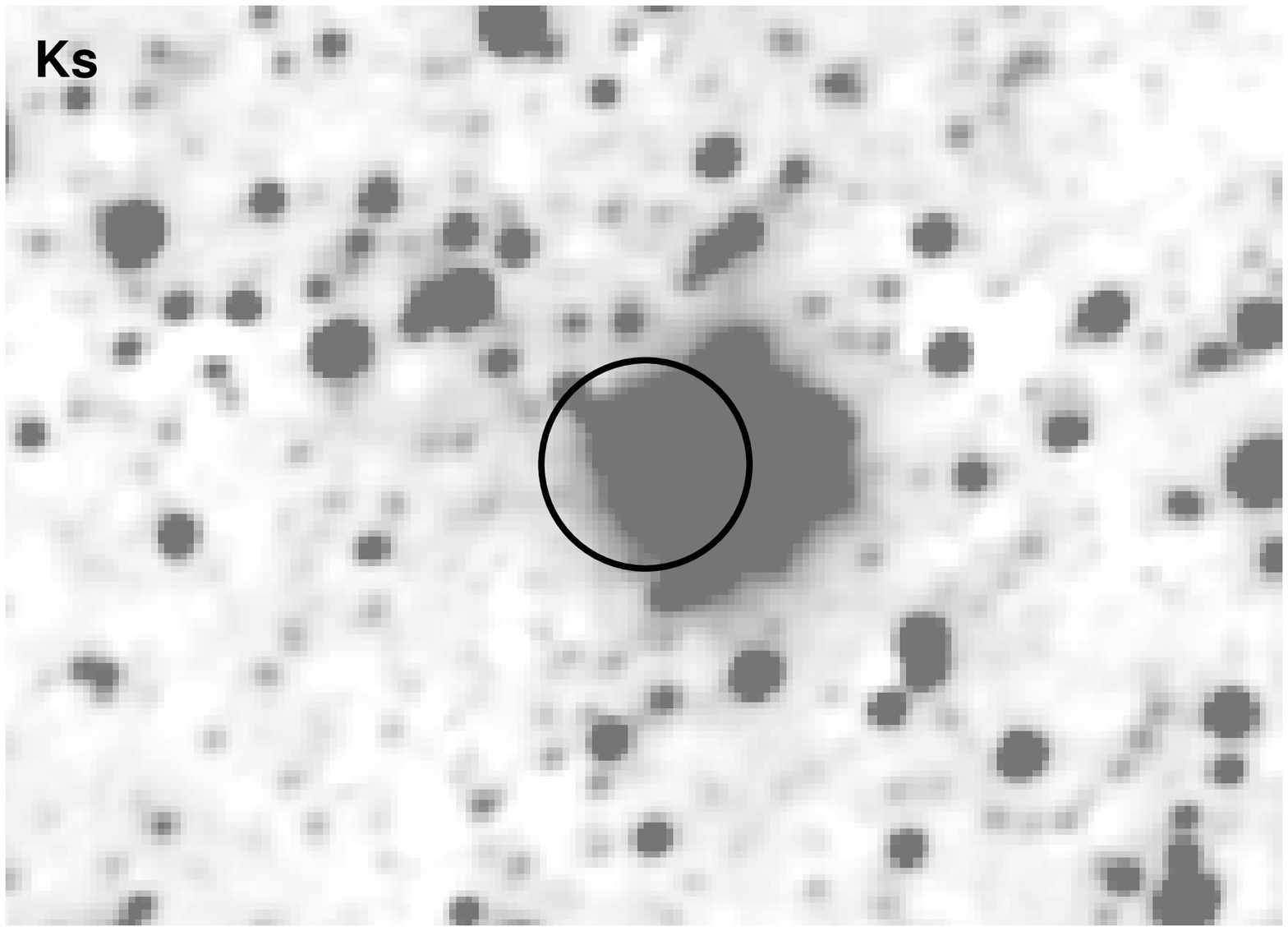}
\caption{The error box of $4\arcsec$ is shown on the $J$ (shortest exposures), $H$ (shortest exposures) and $K_{s}$ frames obtained with SOFI. North is up, east is left. Star labels are given on the $J$ frame.}
\label{pos_star_sofi}
\end{figure}

\begin{figure}
\epsscale{0.3}
\plotone{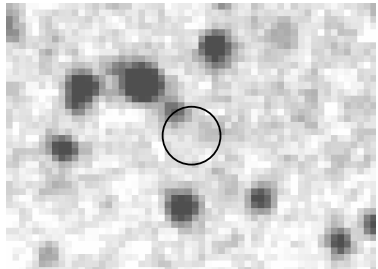}
\epsscale{1}
\caption{The USNO B1.0 plate, IIIaF emulsion, filter RG630. The EPIC error box is reported. North is up, east is left.}
\label{usnofig}
\end{figure}

The EMMI pixel-scale is $0\farcs 166$. Therefore, the error box of $4\arcsec$ reported by EPIC \citep{schartel} corresponds to a circle with a radius of 24 pixels, noticeably greater than the FWHM$\sim 5$ pixels measured on the stars on the processed $BbVRIZ$ frames, and corresponding to the $0\farcs 9$ seeing. To find the pixel coordinates of the center of the EPIC error box, we proceed as follows~:
\begin{itemize}
\item we use the GSC\footnote{\url{http://archive.eso.org/gsc/gsc}} catalog to find the astrometric coordinates of 15 stars in the field with a precision compatible with the EMMI pixel scale~;
\item we use the relevant frame of the DSS catalog\footnote{\url{http://archive.eso.org/dss/dss}} to find visually these stars and mark them on the EMMI frames~; this DSS frame is not actually used for astrometry, only for visual identification~;
\item we compute the transformation matrix to go from the astrometric coordinates to EMMI pixel coordinates~; the residuals are well below the error box, at the $0\farcs 2$ level, compatible with the EMMI pixel-scale and the $0\farcs5$ error reported for GSC\footnote{http://www-gsss.stsci.edu/gsc/gsc2/GSC2home.htm}~;
\item we use this matrix to compute the pixel coordinates of the center of the error box.
\end{itemize}
We show the position of the source with the EPIC error box of $4\arcsec$ superimposed, on EMMI and SOFI frames, respectively on Figure~\ref{pos_star_emmi} and Figure~\ref{pos_star_sofi}.\\
Given the SOFI pixel-scale, the uncertainty of $4\arcsec$ reported by EPIC corresponds to a circle with a radius of 13.9 pixels, noticeably greater than the fwhm$\sim 3$ pixels measured on the stars on the processed $JHK_{s}$ frames. We use a similar method to find the pixel coordinates of the center of the EPIC error box, but we use 2MASS\footnote{\url{http://irsa.ipac.caltech.edu/}} as catalog. Again, the residuals in the construction of the transformation matrix are about a pixel size, e.~g. $0\farcs 3$, with an uncertainty for the 2MASS catalog about $0\farcs09$.\\ 
On Figures~\ref{pos_star_emmi} and~\ref{pos_star_sofi}, the scale is nearly the same, and given by the size of the $4\arcsec$ error box, to allow for visual comparisons. The astrometry is consistent within the errors in both optical and NIR image sets. Three candidates appear within the error box. For further references, we will label them 1, 2 and 3 on Figure~\ref{pos_star_sofi}, left panel. The position of the center of each candidate with respect to the center of the circle is, on frame J~: $3\arcsec$ SW, $3\farcs 6$ NE and $2\farcs 2$ NE, respectively. Figure~\ref{pos_star_sofi} also labels two field stars (4 and 5) for comparisons.\\
We give on Table \ref{astrometry} the astrometry of the three candidates, computed by inverting the transformation matrix from GSC coordinates to pixels on frames $R$, $I$, $J$, and averaging, putting statistical weight according to the uncertainties on the catalogs~; the dispersion being below $0\farcs 1$, and given the uncertainties on GSC and 2MASS catalogs, we take the residual value of $0\farcs5$ as error estimate.
\notetoeditor{It would be far better if figures displayed on~\ref{pos_star_emmi},\ref{pos_star_sofi},\ref{pos_star_sofi},\ref{usnofig} have the same visual scale (given by the black circle, which has always the same physical scale), and fit on consecutive pages, or even on the same page.}
\placefigure{pos_star_emmi}


\placefigure{pos_star_sofi}

\placefigure{usnofig}


\placetable{astrometry}
\begin{deluxetable}{ccc}
\tablecaption{Astrometry of the three candidates, with an uncertainty of $\pm0\farcs2$.\label{astrometry}}
\tablehead{
\colhead{Candidate}&\colhead{$\alpha$}&\colhead{$\delta$}}
\startdata
1&$16^{\rm h}31^{\rm m}48\fs31$&$-48\arcdeg49\arcmin00\farcs7$\\
2&$16^{\rm h}31^{\rm m}48\fs89$&$-48\arcdeg48\arcmin57\farcs7$\\
3&$16^{\rm h}31^{\rm m}48\fs78$&$-48\arcdeg48\arcmin58\farcs7$\\
\enddata
\end{deluxetable}


	\subsection{Broad Band magnitudes of the candidates}
	\label{bandmag}

\begin{figure}
\epsscale{1}
\plotone{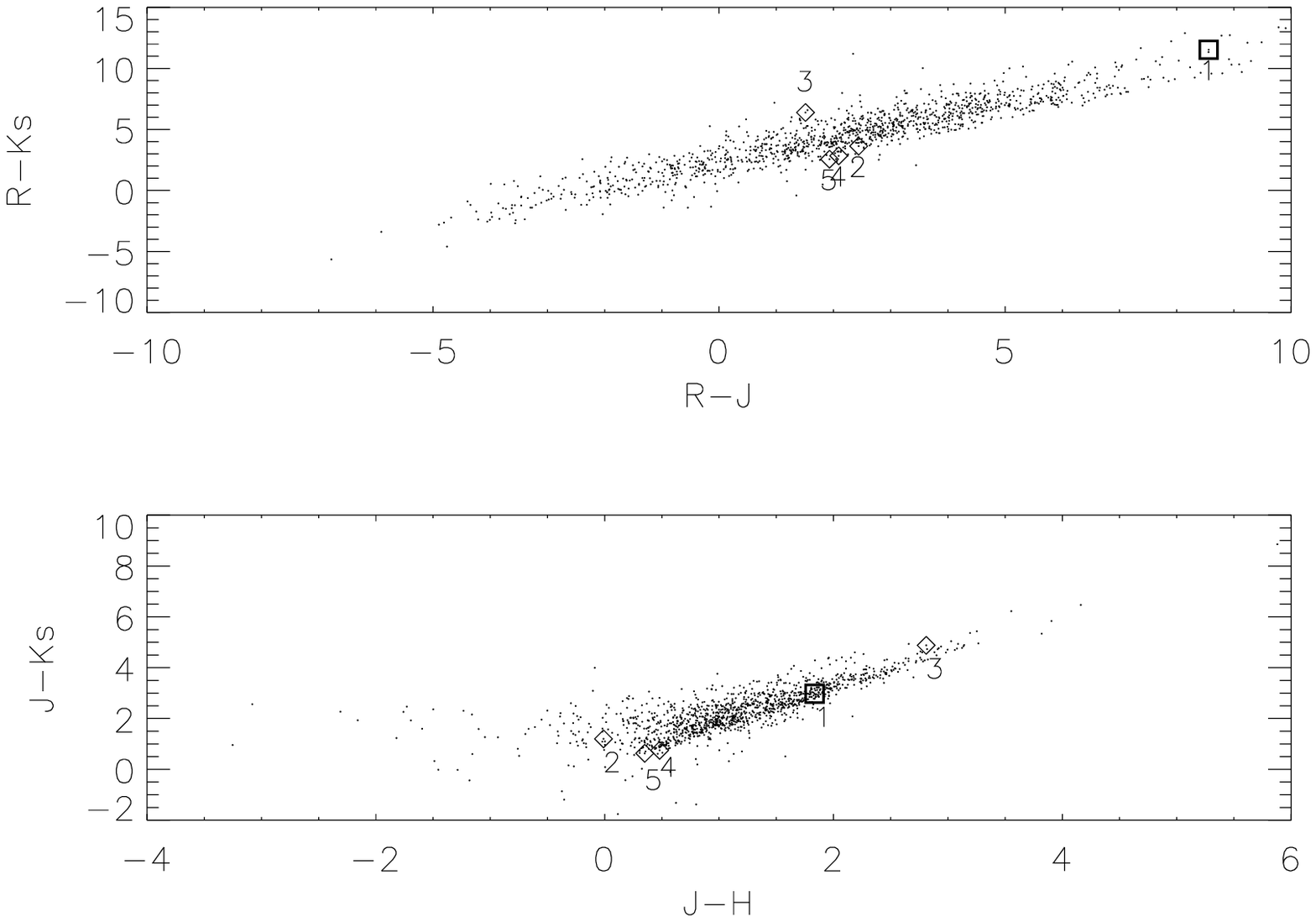}
\epsscale{1}
\caption{Color of stars of the field, detected together in $R$, $J$, $H$, and $K_s$.  2MASS magnitudes were used in $H$ and $K_{s}$ for candidate 1. The numbers 1 to 5 refer to stars lebelled on Figure~\ref{pos_star_sofi}.}
\label{fig_mag}
\end{figure}

We performed photometry in the $BbVRIJHK_{s}$ bands using the {\verb|daophot|} package of {\it IRAF}, for the three candidates reported in Figure~\ref{pos_star_sofi}. For $Bb$ and $V$ frames, we indicate for candidates 1 and 3 the magnitude of the faintest star as an upper limit for the flux. The results are given in Table~\ref{photom_infra_optical}. Table~\ref{photom_archival} is a summary of archival observations of candidate 1.\\
Candidate 2 is barely detectable in $Bb$, more clearly in $V$. Candidates 1 and 3 are detected in the other bands. Candidate 1 is also found in the 2MASS catalog and was suggested as an infrared counterpart by \citet{foschini}~; it is saturating the EMMI detector in $Z$. It is also one of the brightest stars in the NIR field. It saturates in $H$ and $K_{s}$ (even with the shortest exposure time), and in $J$ (shortest exposures) it peaks to 32000 ADU (Analogue to Digital Units), above the limit of linearity of 10000 ADU of SOFI \citep{sofiman}. The position is taken near the maximum, before saturation. Therefore magnitudes for this star reported in Table~\ref{photom_infra_optical} are upper limits in $H$ and $K_{s}$, and there is a possible bias in $J$~; note also that the optical frames show that there is no neighbour in the glare of the star, so it is likely that there is no significant contamination in NIR.\\
Candidate 1 has been seen in the DENIS catalog\footnote{\url{http://cdsweb.u-strasbg.fr/denis.html}}, with magnitudes in agreement with our results. It has been seen in the DSS2\footnote{\url{http://archive.eso.org/dss/dss}} infrared plate (1980), but not seen in the DSS2 red plate (1992). However, as stated by \citet{foschini}, this candidate is not visible on the USNO B1.0 plates\footnote{\url{http://www.nofs.navy.mil/data/fchpix/}} taken with IIIaF emulsion, filter RG630 (see Figure~\ref{usnofig}) and OG 590, corresponding to the $R$ band, obtained respectively in 1982 and 1992. 
This is in contradiction with \citet{walter}, who identified a counterpart in the USNO B1.0 catalog~; however, it appears at the light of our data that their identification corresponds to our candidate 2, whereas their identification with 2MASS is actually our candidate 1. This mistake was probably due to the fact that candidates 1 and 2 have nearly the same magnitude in $R$ in our results, and that candidate 2 is reported to vary of one magnitude over an interval of 50 years (between the two epochs of the USNO catalog). Our observation of candidate 1 in the $R$ band is therefore, up to our knowledge, a discovery. As the magnitude limit on the USNO B1.0 catalog is around 20, our detection seems to imply a variability in $R$ band of more than two magnitudes, whereas there is no variability on $J$ between 2003 (our data) and 1999 (2MASS). Such a strong variability in the optical (although comparisons with photographic measurements of objects of such unusual colors may be rather uncertain, because of slightly different bandpasses), associated to the high variability of IGR~J16318$-$4848 in the X-rays, would suggest that candidate 1 is the genuine counterpart of IGR~J16318$-$4848.\\
We also seek for the source in the IRAS Point source catalog V2.1\footnote{\url{http://irsa.ipac.caltech.edu/IRASdocs/iras.html}}, and find no source closer than $230\arcsec$. We found 1900 IRAS sources in a $3\arcdeg$ radius around the EPIC position. We choose as a robust upper limit for Table~\ref{photom_archival} the flux such as 90\% of the IRAS sources have greater fluxes~: indeed, the flux distribution shows a obvious deficit of fainter sources.\\
Radio observations were performed with the Australia Telescope Compact Array (ATCA) at 4.8 and 8.6 GHz \citep{walter} on 2003, February 9, and show no detection with a $1\,\sigma$ upper limit of 0.1 mJy.\\
The differences in magnitude $R-J$, $R-K_s$, $J-H$, $J-Ks$ of the stars of the field are shown on Figure~\ref{fig_mag}. The unusual colors of candidate 1 are obvious, suggesting a reddening by absorption : the star is within the bulk of the distribution in the $J-H$, $J-K_s$ plane, but is a clear outlier in the $R-J$, $R-K_s$ plane, where absorption is more critical. As being brighter at longer wavelength is expected for a highly obscured source, this suggests that candidates 2 and 3 may be field stars, and candidate 1 is therefore the most likely candidate for IGR~J16318$-$4848 counterpart.\\


\placetable{photom_infra_optical}
\begin{deluxetable}{cccc}
\tablecaption{Photometry of the stars labeled in Figure~\ref{pos_star_sofi}.\label{photom_infra_optical}}
\tablehead{
\colhead{Band}&\colhead{1}&\colhead{2}&\colhead{3}}
\startdata
$Bb$	&$>25.4\pm 1$	&$22.70\pm 0.12$&$>25.4\pm 1$	\\
$V$	&$>21.1\pm 0.1$	&$19.67\pm 0.03$&$>21.1\pm 0.1$	\\
$R$	&$17.72\pm 0.12$&$17.86\pm 0.05$&$20.25\pm 0.14$\\
$I$	&$16.05\pm 0.54$&$17.71\pm 0.06$&$19.74\pm 0.17$\\
\hline
$J$	&$10.33\pm 0.14$&$16.42\pm 0.14$&$16.97\pm 0.14$\\
$H$	&$<10.35\pm 0.15$&$16.43\pm 0.16$&$16.93\pm 0.16$\\
$K_{s}$	&$<9.13\pm 0.10$&$15.22\pm 0.20$&$14.86 \pm 0.17$\\
\enddata
\end{deluxetable}

\placetable{photom_archival}
\begin{deluxetable}{lccc}
\tablecaption{Archival data on candidate 1.\label{photom_archival}}
\tablehead{
\colhead{Name}&\colhead{Year}&\colhead{Band}&\colhead{Result}}
\startdata
DSS2&1980&Infrared plate&seen\\
&1992&Red plate&unseen\\
USNO B1.0&1982&RG630&unseen\\
&1992&OG590&unseen\\
2MASS&1999&$J$&$10.162\pm0.018$\\
&&$H$&$8.328\pm0.037$\\
&&$K_s$&$7.187\pm0.015$\\
DENIS&2003&$I$&$16.217\pm0.06$\\
&&$J$&$10.239\pm0.05$\\
&&$K$&$7.255\pm0.07$\\
IRAS&1983&$12\,\micron$&$< 0.9\,\rm Jy$\\
&&$25\,\micron$&$< 0.7\,\rm Jy$\\
&&$60\,\micron$&$< 4\,\rm Jy$\\
&&$100\,\micron$&$< 40\,\rm Jy$\\
ATCA&2003&8.6 GHz&$< 0.1\,\rm mJy$\\
&&4.8 GHz&$< 0.1\,\rm mJy$\\
\enddata
\end{deluxetable}


\placefigure{fig_mag}


\subsection{Summary}

In the two previous subsections, we have shown that three sources have positions compatible with the error box given by EPIC. Candidate 1 is seen on 2MASS, and we confirm its photometry in the $J$ band. We discovered this candidate in the $R$ band, since it is not detected in the USNO-B1.0 catalog. The dependence of its magnitudes on the wavelength shows a strong brightening at longer wavelengths, which indicates that it is strongly absorbed, whereas the two other candidates have behaviors comparable to the field stars. Therefore, in the rest of this paper, we will consider candidate 1 as the genuine counterpart of IGR~J16318$-$4848.

\section{THE SPECTRAL ENERGY DISTRIBUTION}
\label{secsed}

	\subsection{Spectral continuum}

\begin{figure}
\epsscale{0.8}
\plotone{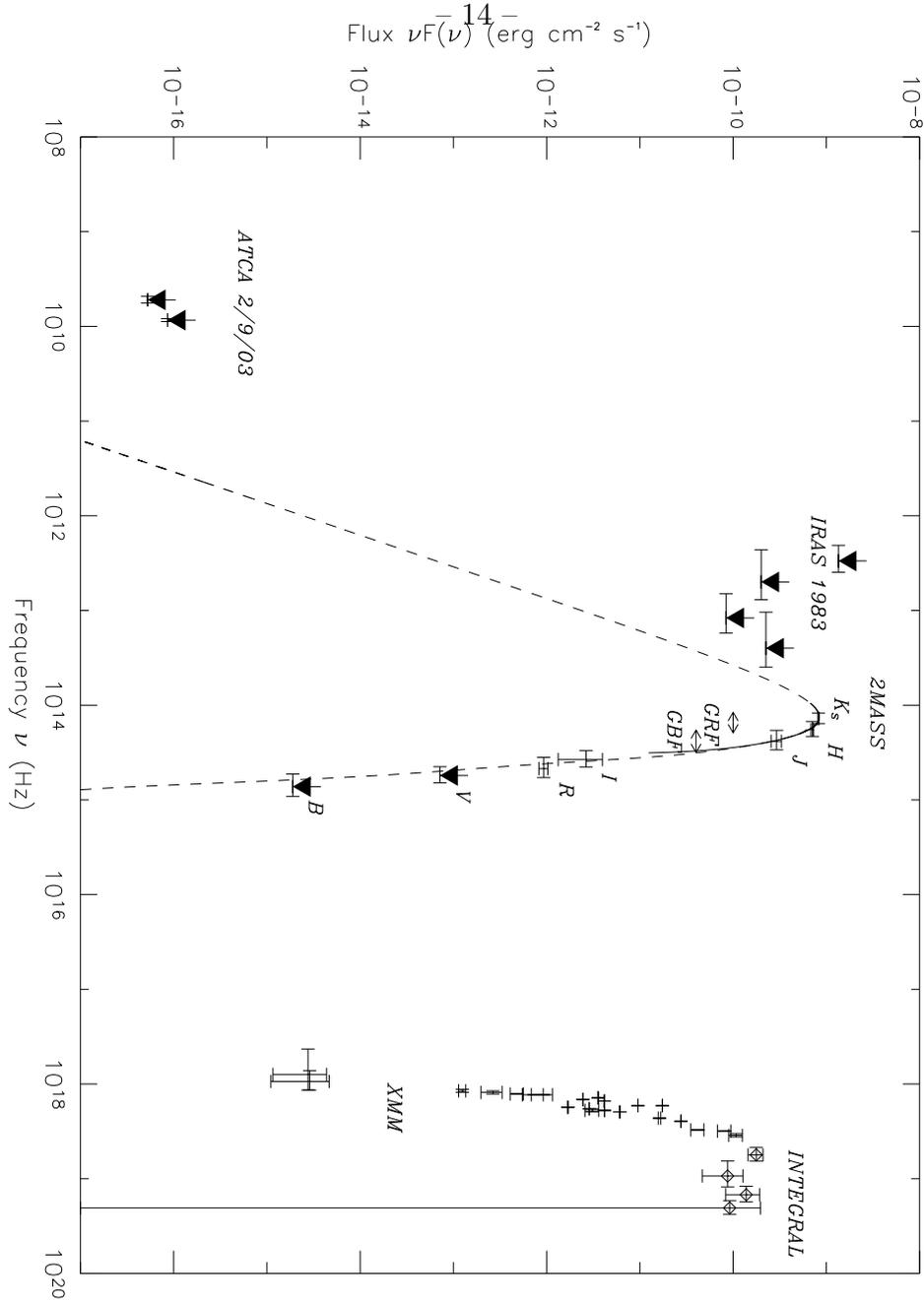}
\caption{Observed SED  in $\left(\nu,\,\nu F(\nu)\right)$ units, including the results of Table~\ref{photom_infra_optical} and the archival data of Table~\ref{photom_archival}. At this stage, the rescaled continuum spectra in GBF and GRF and literature data. The $B$ and $V$ data are upper limit only. The dashed curve corresponds to the fit discussed in section~\ref{discuss_specphot} ~: an absorbed black body, representing well the data. The results of INTEGRAL, XMM, IRAS and ATCA are also shown.}
\label{specphot}
\end{figure}

The continua of both GBF and GRF spectra were extracted, and scaled to match our $J$ magnitude and the 2MASS $K_{s}$ magnitude for candidate 1. This method is approximate, given the unusual colors of the object, nevertheless this leads to a very good agreement in the overlapping region, and with the corresponding 2MASS $H$ magnitude. This approach assumes that the observed magnitudes are mainly due to the continua~; indeed, both in GRF and GBF, the power of the continua represents about 95~\% of the total power. In Figure~\ref{specphot}, we show the Spectral Energy Distribution (SED) in $\left(\nu,\,\nu F(\nu)\right)$ units, including the two continua, with results from Table~\ref{photom_infra_optical} and the archival data of Table~\ref{photom_archival}. At this stage, no dereddening is done. We have also included in the SED a compilation of several observations of the source, covering a very wide range of wavelengths, going from centimeter (ATCA) to X-rays (XMM, INTEGRAL), over 10 decades in wavelength. Data confirming our observations, as for instance the DENIS catalog, are not reported on the SED. However, two caveats have to be mentioned~:
\begin{itemize}
\item the data were taken at different dates, and the source is variable (see \citet{walter}, and our comment on the non-detection of the source on the USNO B1.0 catalog)~;
\item the emissions at different wavelengths can come from different components of the system, as the source is likely to be a complex system involving a compact object, a companion star of an early type, and dust (see below).
\end{itemize}
Therefore, throughout the discussion, we will emphasize on our optical/NIR data, and use the other data mainly for completion or confirmation.

\notetoeditor{This figure should appear on a full single page for clarity.}
\placefigure{specphot}


\placefigure{attracteur}

	\subsection{Absorption and temperature estimates}
	\label{discuss_specphot}

\begin{figure}
\epsscale{0.7}
\plotone{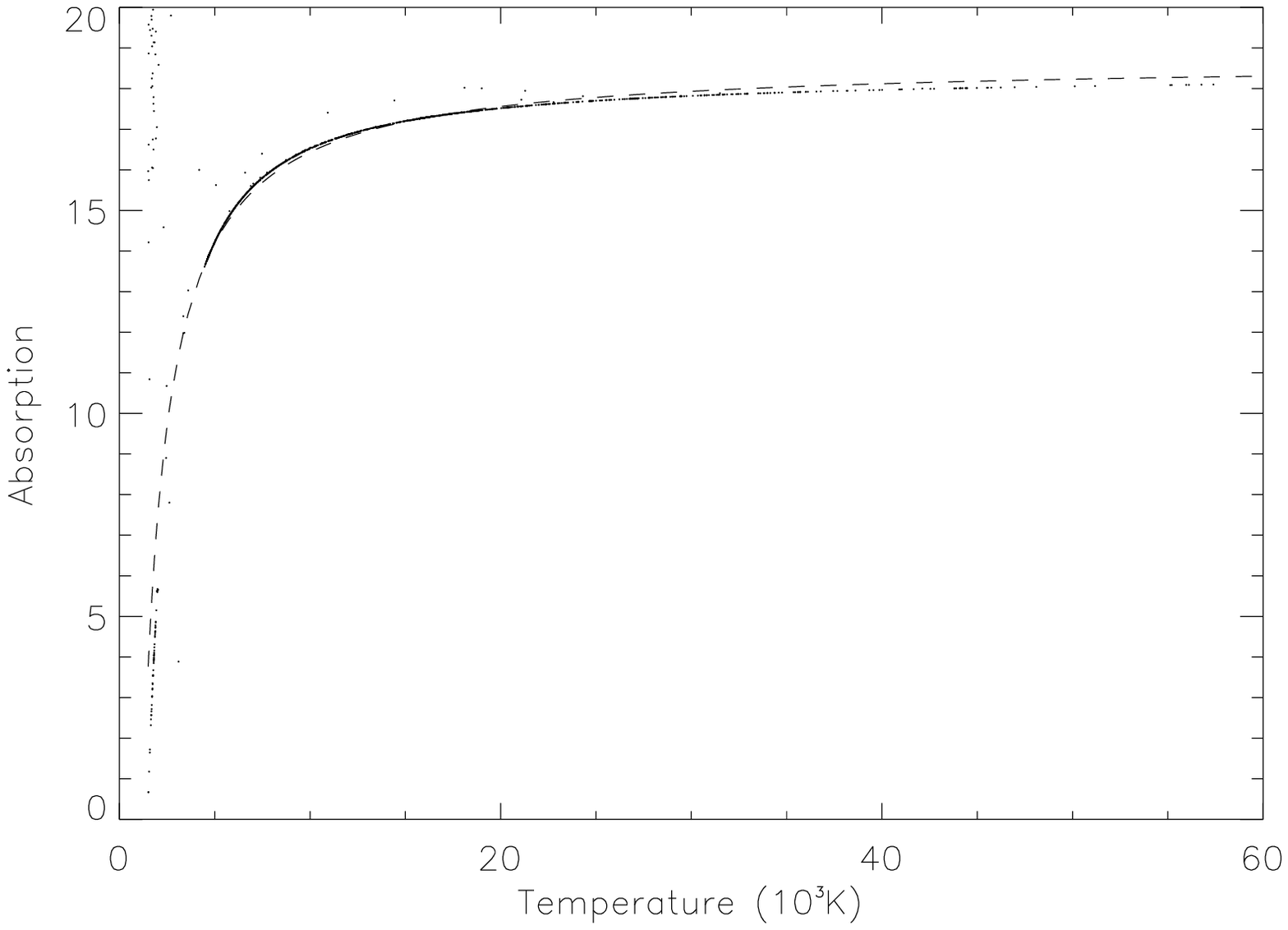}
\plotone{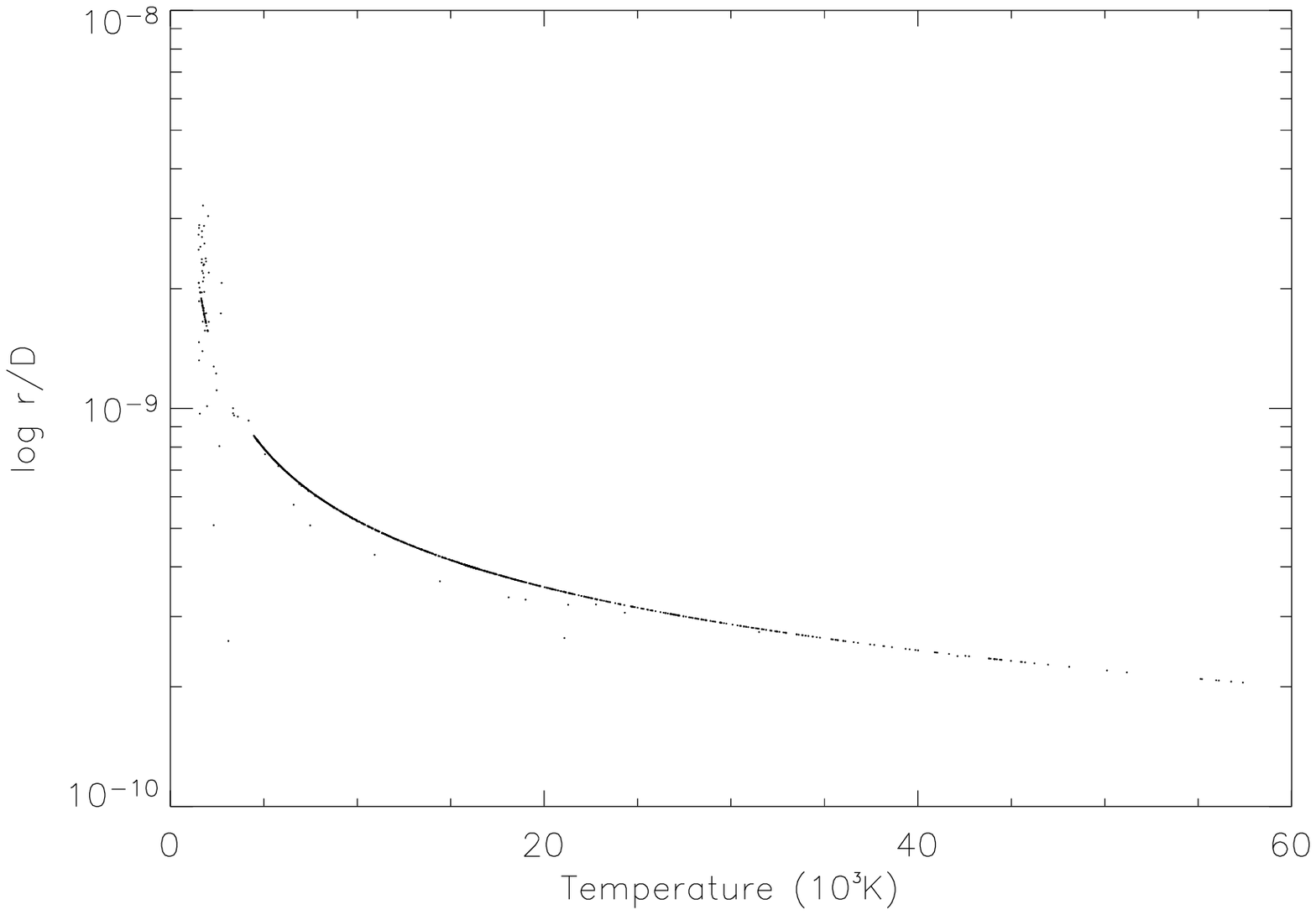}
\epsscale{1}
\caption{The attractor solution for the minimization procedure to fit an absorbed blackbody to data plotted on Figure~\ref{specphot}. Upper plot is for absorption, lower plot is for $r/D$. The points converge towards an attractor, that is fitted by a dash line on the upper plot. The remaining spreaded points indicate that the 1~\% tolerance criterion required for convergence has not been reached in 50 iterations. See text for a discussion of the fit.}
\label{attracteur}
\end{figure}

We suppose that our NIR (photometry and spectroscopy) and optical data are measurements of a blackbody at temperature $T$ with absorption $A_{V}$. If the emitting region is spherical of radius $r$ at a distance $D$ from Earth, then the measured flux in mJy is given by~:
\begin{equation}
f(\nu)=\frac{2\cdot 10^{29}\pi h}{c^{2}}\frac{r^{2}}{D^{2}}\frac{\nu^{3}}{\exp\left(\frac{h\nu}{kT}\right)-1}\cdot A_{\nu}
\end{equation}
where $A_{\nu}$ is the absorption at frequency $\nu$, given $A_{V}$, and is computed according to the formulae of \citet{cardelli}. We fit our $R$, $I$, GBF and GRF fluxes for parameters $r/D$, $T$ and $A_{V}$ using a gradient minimization method. The fit shown on Figure~\ref{specphot} has $r/D=5\cdot10^{-10}$, $T=20\,250\,\rm K$, $A_{V}=17.5$, but because of a strong degeneracy, this is not the sole choice for these parameters.\\ 
This fit reproduces our data well, rendering plausible the hypothesis that an absorbed black body is  the origin of the continuum emission. Note that the absorption law is unknown for wavelengths above $3.3\,\micron$ ($9\times 10^{13}\,\rm Hz$) \citep{cardelli}, and we assume that the expression is still valid for longer wavelengths. The value of the fit is quickly decreasing at frequencies lower than the $K_{s}$ band, rendering likely the non detection of the source by IRAS and ATCA~: at $2.5\times 10^{13}\,\rm Hz$, the average flux of the fit is only 16~\% of the upper limit of IRAS.\\
However, the parameters of the fit are strongly degenerate, and therefore the parameter estimates returned by the minimization method depend strongly on the first guess. Physically, it means that a cool, close source with low absorption can fit the data as well as a hot, distant and absorbed one. By analogy to chaotic mechanic systems, we call ``attractor'' the locus in the parameter space of the various minimum $\chi^{2}$ we can get by varying the first guess. 
To trace this attractor, we use the minimization method 5000 times, with uniformly distributed parameters over $r/D=10^{-12}-10^{-8}$, $T=1\,000-61\,000\,\rm K$, $A_{V}=0-25$. The result is shown on Figure~\ref{attracteur}. There is a clear attractor solution, all individual points corresponding to undistinguishable fits with nearly identical $\chi^{2}$. In the temperature/absorption plane, almost all points converged towards a line well fitted by~:
\begin{equation}
\frac{A_{V}}{A_{V0}}=1-\frac{T_{0}}{T}
\end{equation}
with $A_{V0}=18.68$, $T_{0}=1\,200\,\rm K$. Note that there is a slight discrepancy for high temperatures, indicating that this simple empirical model is not fully satisfying. The $T_{0}$ parameter corresponds to the black body temperature that fit our data if there is no absorption. The $A_{V0}$ parameter is the absorption limit when $T\rightarrow \infty$, i.e. the maximum absorption compatible with our data, and the convergence towards this limit is rather quick. Two regions can be loosely distinguished, corresponding to two different physical situations~:
\begin{description}
\item[a low temperature] (below $\sim 6\,000\,\rm K$) where the absorption is very badly constrained between $A_{V}=5$ and 15. This temperature is compatible with a main sequence dwarf star photosphere, a cool red giant, or even with hot dust if the temperature is not too much above $1000\,\rm K$. On the other hand, the heating of dust can be non thermal (free-free for instance). 
This case corresponds also to a higher $r/D$ ratio, indicating a rather close and/or large source.
\item[a high temperature] (above  $\sim 10\,000\,\rm K$), where the absorption is rather strongly constrained between $A_{V}=16$ and 18.7. This temperature is compatible with an early type star photosphere. However, because the NIR corresponds to the Rayleigh-Jeans region of the spectrum, the temperature has almost no effect on the shape of the spectrum, and is therefore very weakly constrained.
\end{description}
This ambiguity of fitting with an absorbed blackbody had been already noted by \citet{rev1}, using archival observations. We essentially agree with their results.\\
The absorption can be intrinsic, or along the line of sight (interstellar absorption), or both. In the case of interstellar absorption, an upper limit (valid for a source not too close) can be obtained with the $N_{\rm H}$ tool of HEASARC\footnote{\url{http://heasarc.gsfc.nasa.gov/cgi-bin/Tools/w3nh/w3nh.pl}}, using the results of \citet{dickey}. Within $1\arcdeg$ around the position of the source, seven data points are available, well distributed and well centered. By averaging, and putting weight in order to favor closer points of the source, we obtain an estimate of the H {\sc i} column density of $2.11\pm 0.3\,10^{22}\,\rm cm^{-2}$. Using the relation of \citet{predehl}, we get an estimate for the interstellar absorption of $A_{V}=11.8\pm1.6$. If the temperature is above $\sim 10\,000\,\rm K$ (the ``high temperature case'' above), this seems to point to an absorption more than $3\,\sigma$ above interstellar absorption, and hence is an indication of a strong intrinisic absorption, and that, if the star responsible for optical/NIR radiation is this hot, then we see it through a rather dense circumstellar material. This indication is in agreement with the fact that the source, if it is the candidate 1 reported on Figure \ref{fig_mag}, exhibits unusual colors, whereas the neighbour stars labeled 2 to 5 have color not different to those of field stars. However, given the proximity of the source to the Galactic plane, the interstellar medium may be patchy, hence the derived value of interstellar absorption must be taken only as an estimate.\\ 
A more detailed modelization of the SED should include the contribution of the dust, taking into account the distribution of the material, and its physical properties. However, valuable contraints would come from data in the mid-infrared range, where the dust contribution is expected to be noticeable. In the $K_{s}$ band, the contribution of the dust is likely to be hard to distinguish, and indeed our fit is satisfactory in this band. That is the reason why we choose to keep a simple absorbed blackbody model, being not in contradiction with the possible presence of dust.


\section{THE SPECTROSCOPY}
\label{speclines}

In this section, we focus on the spectroscopic lines in order to distinguish between the ``low temperature'' and the ``high temperature'' case of Figure \ref{attracteur}. We will also gather key information to assign a spectral type to the mass donor of the source. 

	\subsection{Line identification}
\label{lines}
	The full GBF and GRF spectra are presented on Figures~\ref{figspec_gbf},~\ref{figspec_grf_part1} and~\ref{figspec_grf_part2}. The GRF spectrum is cut in two parts for a clearer view. The spectra display around 80 lines, most of them being emission lines. All lines selected here are detected above the $2\sigma$ level, and the relative uncertainty in flux is roughly 12~\%.\\
Line fitting is performed with the {\verb|splot|} routine of {\it IRAF\/}, assuming a Gaussian shape.  Tables~\ref{ident_gbf},~\ref{ident_grf_part1} and~\ref{ident_grf_part2} (corresponding to Figures~\ref{figspec_gbf},~\ref{figspec_grf_part1} and~\ref{figspec_grf_part2}, respectively) report the central wavelength for the identification and for the fit, peak flux, equivalent width and FWHM. A negative figure for width indicates that it is an emission feature, a positive figure indicates an absorption feature. To assess if the lines are real features of the object,  we also extracted the spectrum on a different aperture (for GBF only). This spectrum is an order of magnitude below the object spectrum and clearly featureless (except in the strongly absorbed region between $1.35$ and $1.45\,\micron$), proving that the sky subtraction worked successfully in order to remove the sky lines.\\ 
Identification is done using the UKIRT catalog\footnote{\url{http://www.jach.hawaii.edu/JACpublic/UKIRT/astronomy}}, and the works of \citet{morris}, \citet{clark}, \citet{bando}. Only the strongest lines are fitted in the noisiest regions. A question mark indicates that the identification is not certain, or that the fitted value is very sensitive to continuum parameters given in input to {\verb|splot|}, and therefore highly uncertain. Several identifications were often possible, given our resolution. Note that there is always an ambiguity between hydrogen and \ion{He}{2} lines, i.e. a transition \ion{He}{2} (2n-2m) has nearly identical wavelength as the (n-m) hydrogen transition. In any case, there are only two very weak \ion{He}{2} transitions with odd levels (25-8 and 20-9), suggesting that the contribution of even levels to hydrogen lines is negligible. Therefore, this ambiguity is not indicated in the tables.\\ 
In very few cases, there seems to be a double peak structure, the lower peak is then indicated by a ''?-'' in the tables (for instance, the Pa (6-3) at $1.0941\,\micron$  is preceded by a peak at $1.0911\,\micron$). The separation is around $400\,\rm km\,s^{-1}$, similar to the value reported for Ci Cam \citep{clark}, but the result is by far too dubious to lead to any firm conclusion. The FWHM values for the lines of Xenon arcs are 21 and $29\,$\AA, and we found that line broadening is not significant for the well-defined lines. The gap between the wavelength of the identified transitions and the fitted values is consistent within $1\,\sigma$ level with no systemic velocity of the object~: $c\Delta\lambda/\lambda=-110\pm130\,\rm km\, s^{-1}$.\\
Four lines show a P-Cygni profile, indicating that there is a stellar wind of circumstellar material. All four are \ion{He}{1}~: $\rm 2s^{3}S-2p^{3}P^{0}$ ($1.0832\,\micron$), $\rm 3p^{3}P^{0}-5d^{3}D$ ($1.1970\,\micron$), $\rm 2s^{1}S-2p^{1}P^{0}$ ($2.0587\,\micron$), $\rm 3p^{3}P^{0}-4s^{3}S$ ($2.1126\,\micron$). The P-Cygni profile is evident in the two latter cases, more dubious in the former two. Putting more statistical weight on the two last lines, we infer a wind velocity of $410\pm 40\,\rm km\,s^{-1}$, the uncertainty corresponding to the dispersion of the result using the four lines. No other line shows any clear P-Cygni profile.\\
No significant time variation of the lines (in flux or in wavelength) was found, but we cannot give constraints on this because of the short total duration of the observations (10 minutes between the first and the last GBF spectrum).\\
Many spectral lines are common to IGR~J16318$-$4848 and CI Cam \citep{clark}~: 47 lines out of the 57 identified lines of IGR~J16318$-$4848 are seen in CI Cam between 1 and $2.35\,\micron$ \citep{clark}, and conversely 42 over 68 identified lines of CI Cam are seen in IGR~J16318$-$4848~; lines after $2.35\,\micron$  were not listed by \citet{clark}, and the CO bands are not seen in the source, but Pfund lines are common with CI Cam~; an unidentified feature at $1.9855\,\micron$  is common to both sources~; the 2--10 keV X-ray spectrum of IGR~J16318$-$4848 \citep{matt} is very similar to the X-ray spectrum of CI Cam \citep{boirin}. These are hints that the two systems are physically similar.

	\subsection{Hydrogen Lines}

	The Paschen, Brackett and Pfund series are obvious on the spectra. They allow us to possibly estimate the absorption, using the Brackett decrement, independently of the SED fitting. The relative intensities of the hydrogen recombination lines have been calculated by \citet{hummer} for an electron temperature of $10^{4}\,\rm K$ and an electron density of $10^{4}\,\rm cm^{-3}$ and considering Case B recombination. We compute the integrated fluxes of our Brackett and Paschen lines and divide by these relative intensities to obtain the absorption law. We fit separately the Brackett GBF lines and the Brackett GRF lines (all the lines) for the absorption $A_{V}$, and put a very low statistical weight on lines for which the intensity is uncertain, e.g. those at the edges of the GBF and GRF spectra, or within the noisiest regions. Results are given in Figure~\ref {decbrackett}. GBF and GRF lines, considered separately, give somewhat different absorption values, 25.4 and 18.1 ; the joint fit gives a value of $A_{V}=18.2$. However, in the fitting procedure, the minimum value of absorption is not very sharply defined. Note also that Br (7-4) (or Br$\gamma$) has a smaller flux than expected with respect to the higher order Brackett lines, suggesting that the lines are optically thick or self-absorbed. Nevertheless, this method clearly favors the high temperature part of Figure~\ref{attracteur}, because it indicates that absorption is higher than 16. Therefore, it is likely that we are actually observing an early type hot star through a dense absorbing column, rather than a cool late star with lower absorption.

	\subsection{Helium Lines}

	The upper state of \ion{He}{1} $1.0832\,\micron$  doublet, a strong line on our spectrum,  can be populated by recombination, or can be collisionally excited from the metastable $\rm 2s^{3}S$ state, hence its flux can be large even when the ionized fraction of helium is low \citep{gregor}. Therefore, this line can come either from ionized or neutral regions. The $2.0581\,\micron$  transition is among the strongest in our spectrum : this could be a hint of the presence of a high-density region close to the star, in this case the $584\,\rm\AA$  transition would be optically thick \citep{gregor}, but because of the strong absorption, this transition will be very difficult to observe. Because two mechanisms can populate these transitions, the  $1.0832\,\micron$ and $2.0581\,\micron$ lines are not reliable indicators of circumstellar conditions.\\
We may estimate abundance of helium from the intensity ratios of \ion{He}{1} $1.197\,\micron$  and $1.7007\,\micron$ to Br (7-4). This is a lower limit for the helium abundance, because as the ionization energy of helium is greater than the one of hydrogen we have $N(\rm He^{+})/N(\rm H^{+}) < N(\rm He)/N(\rm H)$, and because \ion{He}{1} will not arise in the shielded $\rm Fe^{+}$ zones. According to \citet{allen}, we have~:
\begin{equation}
\frac{N(\rm He^{+})}{N(\rm H^{+})}=0.79\frac{I(1.7007\,\micron)}{I(\rm Br (7-4))}=1.15\frac{I(1.197\,\micron)}{I(\rm Br (7-4))}
\end{equation}
The result depends on the absorption $A_{V}$ we choose to deredden the lines, as the line wavelengths are remote from each other. The impact of absorption on the second ratio is higher, because of a greater distance between the two wavelengths, and because the line at $1.197\,\micron$ has an absorbed flux much weaker than the one at $1.7007\,\micron$, there must exist an absorption value for which the two ratios are equal. This happens for $A_{V}=17.4\pm 2.5$, the uncertainty coming from a relative uncertainty of 4~\% on the flux lines. It is remarkable that this value is within the attractor values in the high temperature region of Figure~\ref{attracteur} (the value of 17.4 is associated with a temperature of $18\,000\,\rm K$, however the constraint on temperature is very loose, as it can be seen in Figure~\ref{attracteur}). The corresponding value is $N(\rm He^{+}/\rm H^{+})=0.32\pm 0.04$. The first ratio is less subject to absorption, and gives $N(\rm He^{+}/\rm H^{+})$ values ranging from 0.2 to 0.5 for $A_{V}$ going from 10 to 25. In any case, this is a value far higher than the solar value of $N(\rm He)/N(\rm H)=0.067$, indicating an evolved star. Note however that Br (7-4) is smaller than expected considering the higher order Brackett series, and that its intensity, according to Figure~\ref{decbrackett} may be underestimated up to a factor 10, causing an equivalent overestimation of the $N(\rm He^{+}/\rm H^{+})$ ratio.\\
We also consider the relative intensities of lines at $1.0832\,\micron$ (doublet), $1.1970\,\micron$, $1.7007\,\micron$  and $2.0587\,\micron$, and compare to predicted recombination intensities (see e.g. \citet{allen}). The results are given in Table~\ref{heliumallen}. For dereddening, we take $A_{V}=17.4$. Note the concordance (better than 1~\%) between the $1.1970\,\micron$  and $1.7007\,\micron$, another indication that dereddening is correct. We are in a case very similar to the one of $\eta$ Carinae, discussed in \citet{allen}, with a line at $1.0832\,\micron$  about 4 times weaker than expected, and the prominence of the line at $2.0587\,\micron$. Very interestingly, the ratios for CI Cam show a similar behavior \citep{clark}, with a line at $1.0832\,\micron$  about 3 times weaker than expected, and with a $2.0587\,\micron$ line 7 times relatively more prominent than in our data. In the case of $\eta$ Carinae, these facts were explained by invoking a large optical depth for $1.0832\,\micron$, and a significant optical thickness in the $\rm 1^{1}S-1^{1}P$ $584.3\,\micron$  transition of He for $2.0587\,\micron$.\\
Two \ion{He}{2} transitions, at 1.6241 and $1.7717\,\micron$, are present, although very weak. The first one was also seen in CI Cam \citep{clark}. They indicate a high temperature able to ionize helium. Rough flux ratio with \ion{Fe}{2} lines, using the solar abundances, leads to flux of photons able to ionize helium and iron in agreement with a temperature between $10^{4}$ and $2\times 10^{4}\,\rm K$.\\

	\subsection{Metallic Lines}

Several \ion{Fe}{2} lines can be seen in our spectra. Among the strongest, $1.1126\,\micron$  and $2.089\,\micron$  are also seen in early-type high luminosity stars \citep{gregor}, and arise in highly excited levels which are probably populated by UV fluorescence.\\
A  peculiar emphasis must be put on the presence of six forbidden iron lines. Although they are rather weak, the $1.1567\,\micron$  line being for instance  weaker than its neighbors of \ion{He}{1} and N {\sc i}, they provide a very important clue for the identification of the star.\\
The doublet emission of \ion{Na}{1} $\rm 4s^{2}S-4p^{2}P^{0}$ is seen, although not separated. The emission is most likely fluorescent, being pumped by $0.3303\,\micron$  photons. The ionization potential is low (5.1 eV), so that sodium will be largely ionized around the star. Therefore, the sodium emission comes from regions that are not exposed directly to the radiation of the star or of the compact object. This suggests the presence of matter surrounding both objects. The other lines of \ion{Na}{1}, such as 1.1385 and $1.1407\,\micron$, are not seen.\\
The \ion{Mg}{2} doublet at 2.138 and $2.144\,\micron$  is clearly seen. It can be excited by L$\beta$ fluorescence. In our case, the $2.138\,\micron$  line is twice more intense than the $2.144\,\micron$ line,  in agreement with the higher statistical weight of the $\rm 5P_{3/2}$ transition, and also to its shorter distance to L$\beta$.

	\subsection{Molecular Lines}

No lines from $\rm H_{2}$ have been found, suggesting either that shock heating is relatively unimportant, or that temperature is sufficiently high to dissociate the molecules. Molecular hydrogen has not been seen either in CI Cam \citep{clark}, nor in the massive stars in transition studied by \citet{morris}. Emission lines of $\rm H_{2}$ are more usual in massive young stellar objects \citep{kumar}, suggesting that there is no such radiation-protected medium in the environment of this source.\\ 
No $\rm CO$ bands have been found, although a strong 8 bin rebinning of the GRF spectrum shows a step of $\sim 10\,\rm mJy$ over $0.1\,\micron$ at $2.295\,\micron$, locus of the (2-0) bandhead. However, this is too weak and too uncertain. The $\rm CO$ flux, if present, is therefore lower than $\sim 10\,\rm mJy$ (with no dereddening).

	\subsection{Summary}

This study on the spectroscopic lines has shown that there is a dense circumstellar material around the source, possibly enshrouding both the mass donor and the compact object, and with a stellar wind component. This points to a high absorption, and we have seen that the value of $A_{V}=17.4$ is favored from the data. The line ratios favor a high temperature~; along with the high absorption, this is consistent with the case ''high temperature'' of Figure~\ref{attracteur},  and points to a hot early-type star observed through a dense absorbing material. Moreover, the presence of forbidden lines and the fact that almost all the lines are in emission point towards a ``[e]'' star, e.g. a B[e], with a striking similarity with the companion star in CI Cam.

\begin{figure}
\epsscale{0.8}
\plotone{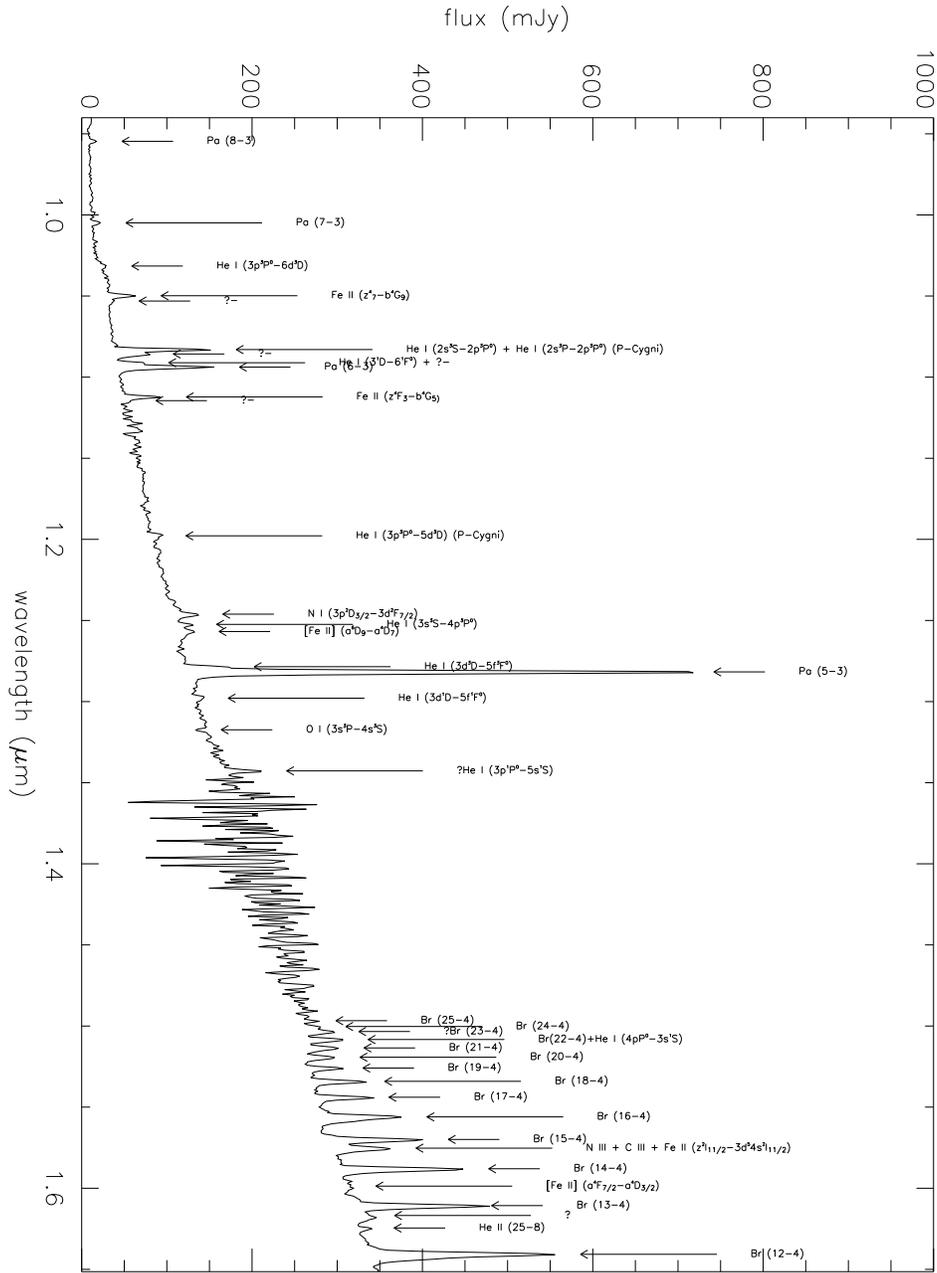}
\caption{The GBF spectrum for the object. Indicated lines refer to Table~\ref{ident_gbf}.}
\label{figspec_gbf}
\end{figure}

\notetoeditor{This figure should appear on a full single page for clarity.}
\placefigure{figspec_gbf}


\begin{figure}
\epsscale{0.8}
\plotone{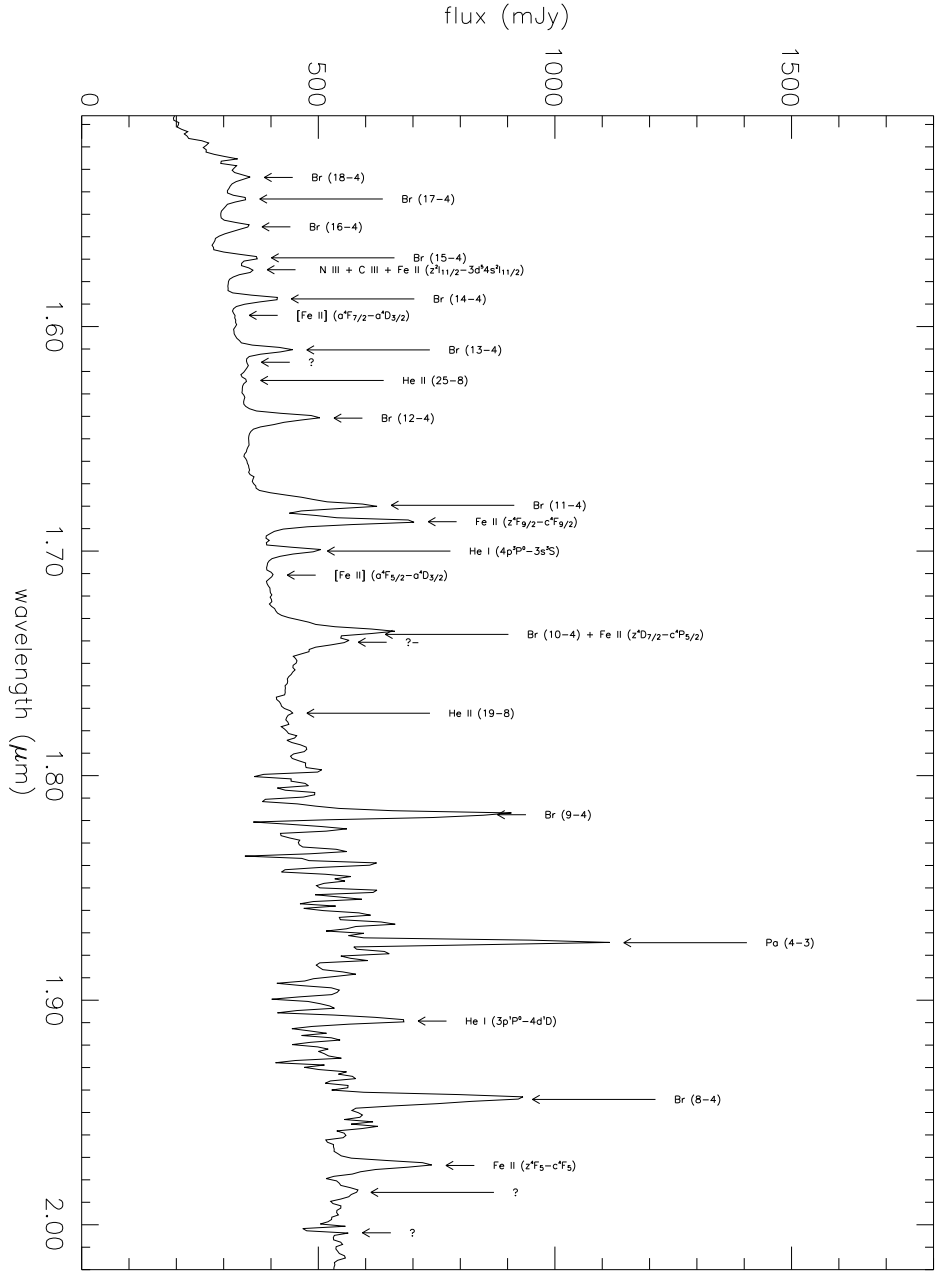}
\caption{The GRF spectrum for the object, first part. Indicated lines refer to Table~\ref{ident_grf_part1}.}
\label{figspec_grf_part1}
\end{figure}

\notetoeditor{This figure should appear on a full single page for clarity.}
\placefigure{figspec_grf_part1}

\begin{figure}
\epsscale{0.8}
\plotone{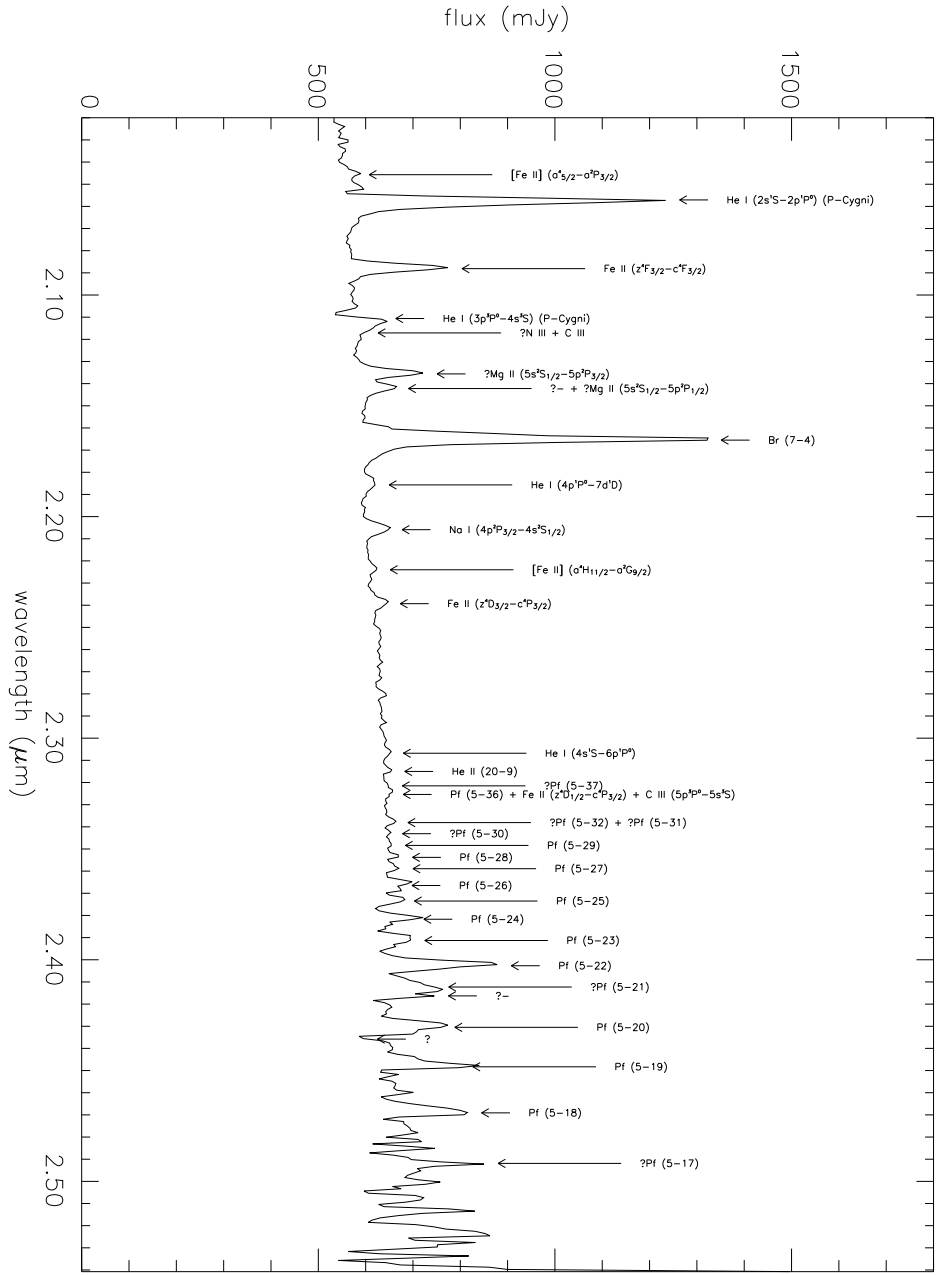}
\caption{The GRF spectrum for the object, second part. Indicated lines refer to Table~\ref{ident_grf_part2}.}
\label{figspec_grf_part2}
\end{figure}

\notetoeditor{This figure should appear on a full single page for clarity.}
\placefigure{figspec_grf_part2}


\placetable{ident_gbf}
\begin{deluxetable}{lccccc}
\tablecaption{Identification of lines on the GBF spectrum.\label{ident_gbf}}
\tablehead{
\colhead{Identification}&\colhead{$\lambda$}&\colhead{$\lambda_{fit}$}&\colhead{Flux}&\colhead{EW}&\colhead{FWHM}\\
\colhead{}&\colhead{\micron}&\colhead{\micron}&\colhead{mJy}&\colhead{\AA}&\colhead{\AA}}
\startdata
Pa (8-3)& 0.9549& 0.9546& 9& -21& 19\\
Pa (7-3)&1.0052	&1.0049& 10& -15& 18\\
\ion{He}{1} (74)($3\rm p^{3}P^{0}-6d^{3}D$)&10314	&1.0315& 4& -2& 15\\
\ion{Fe}{2} ($\rm z^{4}F_{7}-b^{4}G_{9}$)&1.0501	&1.0498& 34& -24& 20\\
?-	&?-	& 1.0531& 6& -6& 30\\
\ion{He}{1} ($\rm 2s^{3}S-2p^{3}P^{0}$)	&1.0832	& 1.0830& 119& -47& 15\\
+\ion{He}{1} ($\rm 2s^{3}P-2p^{3}P^{0}$)	&1.0833 &&&&\\
?-	&?-	& 1.0857& 40& -30& 29\\
\ion{He}{1} ($\rm 3^{1}D-6^{1}F^{0}$)&1.0917&1.0911& 34& -10& 12\\
+ ?- & ?- &1.0911& 34& -10& 12\\
Pa (6-3)&1.0941&1.0938	& 110& -59& 22\\
\ion{Fe}{2} ($\rm z^{4}F_{3}-b^{4}G_{5}$)&1.1126&1.1122& 46& -17& 17\\
?-	&?-	& 1.1145& 7& -3& 19\\
\ion{He}{1} ($\rm 3p^{3}P^{0}-5d^{3}D$)&1.1970&1.1970& 10& -3& 27\\
N {\sc i} (36)($\rm 3p^{2}D_{3/2}-3d^{2}F_{7/2}$)&1.2469&1.2461& 25& -11& 43\\
\ion{He}{1} ($\rm 3s^{3}S-4p^{3}P^{0}$)&1.2528& 1.2524& 19& -3& 18\\
$\left[\right.$\ion{Fe}{2}$\left.\right]$ ($\rm a^{6}D_{9}-a^{4}D_{7}$)&1.2567& 1.2568& 17& -7& 47\\
\ion{He}{1} ($\rm 3d^{3}D-5f^{3}F^{0}$)&1.2789& 1.2785& 49& ?-6& 14\\
Pa (5-3)&1.2822&1.2817& 634& -117& 22\\
\ion{He}{1} ($\rm 3d^{1}D-5f^{1}F^{0}$)&1.2976& 1.2979& 7& ?-1& 18\\
\ion{O}{1} ($\rm 3s^{3}P-4s^{3}S$)&1.3165& 1.3174& -15& 2& 20\\
?\ion{He}{1} ($\rm 3p^{1}P^{0}-5s^{1}S$)&1.3415&1.3427& 48& -6& 21\\
Br (25-4)&1.4971& 1.4967& 21& -1& 14\\
Br (24-4)&1.5005& 1.5003& 21& -2& 21\\
?Br (23-4)&1.5043& 1.5034& 38& -4& 26\\
Br (22-4)&1.5087&1.5083& 49& -5& 25\\
+\ion{He}{1} ($\rm 4p^{1}P^{0}-3s^{1}S$)&1.5088&&&&\\
Br (21-4)&1.5137&1.5135& 40& -5& 29\\
Br (20-4)&1.5196&1.5192& 36& -4& 25\\
Br (19-4)&1.5265&1.5259& 38& -16& 21\\
Br (18-4)&1.5346&1.5341& 65& -6& 23\\
Br (17-4)&1.5443&1.5439& 68& -6& 22\\
Br (16-4)&1.5561&1.5560& 87& -12& 36\\
Br (15-4)&1.5705&1.5699& 110& -10& 25\\
\ion{N}{3}+\ion{C}{3} 13-9&1.575&1.5753& 65& -8& 33\\
+\ion{Fe}{2} ($\rm z^{2}I_{11/2}-3d^{5}4s^{2}I_{11/2}$)&1.576&&&&\\
Br (14-4)&1.5885&1.5880& 147& -14& 27\\
$\left[\right.$\ion{Fe}{2}$\left.\right]$ ($\rm a^{4}F_{7/2}-a^{4}D_{3/2}$)&1.598& 1.5987& 5& -1& 28\\
Br (13-4)&1.6114&1.6107& 158& -14& 27\\
?	&?	& 1.6168& 18& -3& 51\\
\ion{He}{2} (25-8)&1.6241& 1.6245& 12& -1& 13\\
Br (12-4)&1.6412&1.6407& 203& -23& 37\\
\enddata
\end{deluxetable}

\placetable{ident_grf_part1}
\begin{deluxetable}{lccccc}
\tablecaption{Identification of lines on the GRF spectrum, part 1.\label{ident_grf_part1}}
\tablehead{
\colhead{Identification}&\colhead{$\lambda$}&\colhead{$\lambda_{fit}$}&\colhead{Flux}&\colhead{EW}&\colhead{FWHM}\\
\colhead{}&\colhead{\micron}&\colhead{\micron}&\colhead{mJy}&\colhead{\AA}&\colhead{\AA}}
\startdata
Br (18-4)&1.5346& 1.5336& 39& -4& 26\\
Br (17-4)&1.5443&1.5432& 47& -4& 25\\
Br (16-4)&1.5561&1.5556& 66& -10& 40\\
Br (15-4)&1.5705&1.5694& 79& -8& 27\\
\ion{N}{3}+\ion{C}{3} 13-9&1.575&1.5747& 59& -10& 48\\
+\ion{Fe}{2} ($\rm z^{2}I^{0}_{11/2}-3d^{5}4s^{2}I_{11/2}$)&1.576&&&&\\
Br (14-4)&1.5885&1.5877& 108& -11& 30\\
$\left[\right.$\ion{Fe}{2}$\left.\right]$ ($\rm a^{4}F_{7/2}-a^{4}D_{3/2}$)&1.598& 1.5950& 9& ?-6& ?58\\
Br (13-4)&1.6114&1.6104& 118& -12& 32\\
?	&?	& 1.6159& 21& -5& 70\\
\ion{He}{2} (25-8)&1.6241&1.6240& 9& -3?& 85?\\
Br (12-4)&1.6412&1.6408& 150& -21& 44\\
Br (11-4)&1.6811& 1.6796& 217& -33& 52\\
\ion{Fe}{2} ($\rm z^{4}F_{9/2}-c^{4}F_{9/2}$)&1.688&1.6869& 328& -32& 34\\
\ion{He}{1} ($\rm 4p^{3}P^{0}-3s^{3}S$)&1.7007&1.7000& 118& -9& 29\\
$\left[\right.$\ion{Fe}{2}$\left.\right]$ ($\rm a^{4}F_{5/2}-a^{4}D_{3/2}$)&1.711& 1.7107& ?15& ?-1& ?27\\
Br (10-4)&1.7367&1.7371& 179& -45& 90\\
+\ion{Fe}{2} ($\rm z^{4}D_{7/2}-c^{4}P_{5/2}$)&&&&&\\
?-	&?-	& ?17406& ?160& ?-15& ?48\\
\ion{He}{2} (19-8)&1.7717&1.7722& ?17& ?-2& ?32\\
Br (9-4)&1.8181&1.8174& 473& -36& 30\\
Pa (4-3)&1.8756& 1.8744& 597& -22& 20\\
\ion{He}{1} ($\rm 3p^{1}P^{0}-4d^{1}D$)&1.9094& 1.9093& 247& -18& 31\\
Br (8-4)&1.9451& 1.9442& 386& -29& 39\\
\ion{Fe}{2} ($\rm z^{4}F_{5}-c^{4}F_{5}$)&1.9746& 1.9736& 196& -14& 36\\
?	&?	& 1.9856& 42& -3& 35\\
?	&?	& 2.0026& -156& 2& 8\\
\enddata
\end{deluxetable}



\placetable{ident_grf_part2}
\begin{deluxetable}{lccccc}
\tablecaption{Identification of lines on the GRF spectrum, part 2.\label{ident_grf_part2}}
\tablehead{
\colhead{Identification}&\colhead{$\lambda$}&\colhead{$\lambda_{fit}$}&\colhead{Flux}&\colhead{EW}&\colhead{FWHM}\\
\colhead{}&\colhead{\micron}&\colhead{\micron}&\colhead{mJy}&\colhead{\AA}&\colhead{\AA}}
\startdata
$\left[\right.$\ion{Fe}{2}$\left.\right]$ ($\rm a^{4}_{5/2}-a^{2}P_{3/2}$)&2.046& 2.0457& 24& -1& 26\\
\ion{He}{1} ($\rm 2s^{1}S-2p^{1}P^{0}$)&2.0587&2.0581& 628& -42& 35\\
\ion{Fe}{2} ($\rm z^{4}F_{3/2}-c^{4}F_{3/2}$)&2.089& 2.0881& 201& -13& 36\\
\ion{He}{1} ($\rm 3p^{3}P^{0}-4s^{3}S$)&2.1126&2.1116& 63& -5& 43\\
?\ion{N}{3} + \ion{C}{3}&2.116&?2.1151&? 42&? -4&? 44\\
?\ion{Mg}{2} ($\rm 5s^{2}S_{1/2}-5p^{2}P^{0}_{3/2}$)&2.138&2.1356& 135& -8& 35\\
?-	&?-	& 2.1422& 71& -6& 47\\
+?\ion{Mg}{2} ($\rm 5s^{2}S_{1/2}-5p^{2}P^{0}_{1/2}$)&2.144&&&&\\
Br (7-4)&2.1661& 2.1655& 698& -45& 36\\
\ion{He}{1} ($\rm 4p^{1}P^{0}-7d^{1}D$)	&2.1847	& ?2.1857& ?55& ?-7& 50\\
\ion{Na}{1} ($\rm 4p^{2}P^{0}_{3/2}-4s^{2}S_{1/2}$)&2.2056,2.209& 2.2059& 45& -3& 40\\
$\left[\right.$\ion{Fe}{2}$\left.\right]$ ($\rm a^{4}H_{11/2}-a^{2}G_{9/2}$)&2.224& 2.2240& 12& -0& 21\\
\ion{Fe}{2} ($\rm z^{4}D_{3/2}-c^{4}P_{3/2}$)&2.240& 2.2393& 30& -2& 40\\
\ion{He}{1} ($\rm 4s^{1}S-6p^{1}P^{0}$)&2.3069& 2.3068& 17& -1& 40\\
\ion{He}{2} (20-9)&2.314	& 2.3150& 20& -1& 22\\
?Pf (5-37)&2.3218& ?2.3215& ?8& ?-0& ?12\\
Pf (5-36)&2.3242	& 2.3244& 15& -1& 35\\
+\ion{Fe}{2} ($\rm z^{4}D^{0}_{1/2}-c^{4}P_{3/2}$)&2.3247&&&&\\
+\ion{C}{3} ($\rm 5p^{3}P^{0}-5s^{3}S$)&2.324&&&&\\
?Pf (5-32)	&2.3365	& 2.3381& 25& -1& 29\\
+?Pf (5-31)&2.3404&&&&\\
?Pf (5-30)	&2.3445& 2.3431& 14& ?-1& 19\\
Pf (5-29)	&2.3492& 2.3484& 14& -1& 50\\
Pf (5-28)	&2.3545	& 2.3538& 35& -1& 20\\
Pf (5-27)	&2.3604	& 2.3589& 30& -2& 33\\
Pf (5-26)	&2.3669	& 2.3665& 50& -4& 49\\
Pf (5-25)	&2.3744	& 2.3735& 47& -2& 27\\
Pf (5-24)	&2.3828	& 2.3817& 81& -3& 22\\
Pf (5-23)	&2.3925	& 2.3913& 319& -19& 36\\
Pf (5-22)	&2.4036	& 2.4027& 231& -14& 36\\
?Pf (5-21)	&2.4164	& 2.4123& 67& -6& 53\\
?-	&?-	& 2.4153& 65& -4& 39\\
Pf (5-20)	&2.4314& 2.4305& 115& -7& 39\\
?	&?	& 2.4358& -114& 3& 14\\
Pf (5-19)	&2.4490	& 2.4483& 182& -7& 24\\
Pf (5-18)	&2.4670	& 2.4691& 152& -8& 34\\
?Pf (5-17)	&2.4953	& 2.4929& 172& -4& 17\\
\enddata
\end{deluxetable}

\begin{figure}
\plotone{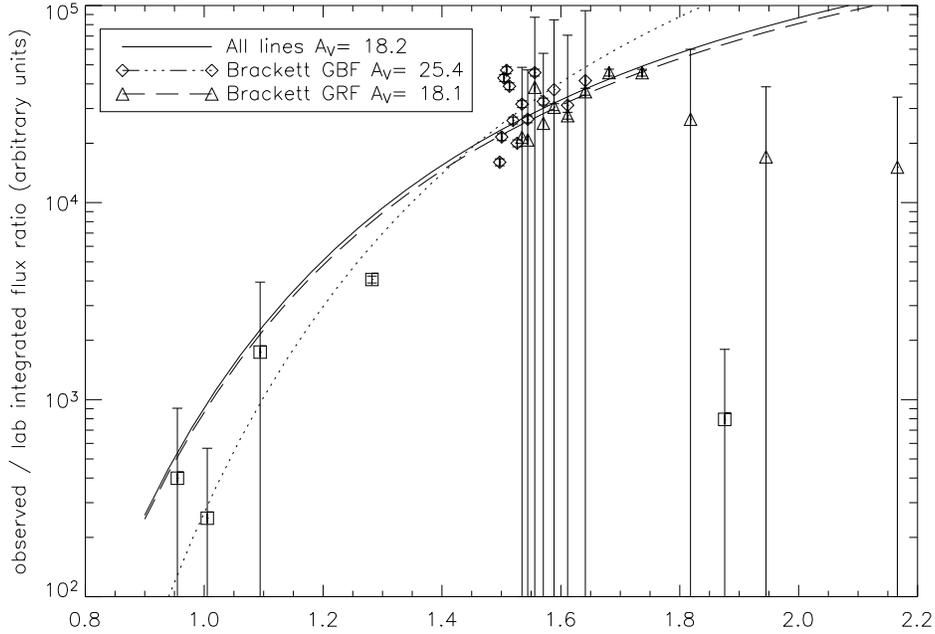}
\caption{The observed integrated fluxes of hydrogen lines divided by the relative intensities from \citet{hummer}. The lines indicate the best fit absorption model for the Brackett lines in the GBF spectrum (data in diamonds, fit in dots), in the GRF spectrum (triangles, dashes), all lines including Paschen (squares, solid line). Error bars are 4~\% relative in flux, except for the lines at the edges and in the noisiest regions.}
\label{decbrackett}
\end{figure}

\placefigure{decbrackett}


\placetable{heliumallen}
\begin{deluxetable}{cccc}
\tablecaption{Helium lines comparisons.\label{heliumallen}}
\tablehead{
\colhead{Wavelength}&\colhead{Dereddened intensity}&\colhead{Predicted intensity}&\colhead{Observed/Predicted}\\
\colhead{\micron}&\colhead{mJy}&\colhead{$1.700\,\micron$=1}&\colhead{$1.700\,\micron$=1}}
\startdata
1.0832	&35280	&71.4	&0.27\\
1.1970	&1274	&0.69	&1\\
1.7007	&1852	&1	&1\\
2.0587	&4754	&0.01	&257\\
\enddata
\end{deluxetable}


\section{DISCUSSION}
\label{secdiscuss}

	\subsection{Distance Estimate}

\begin{figure}
\plotone{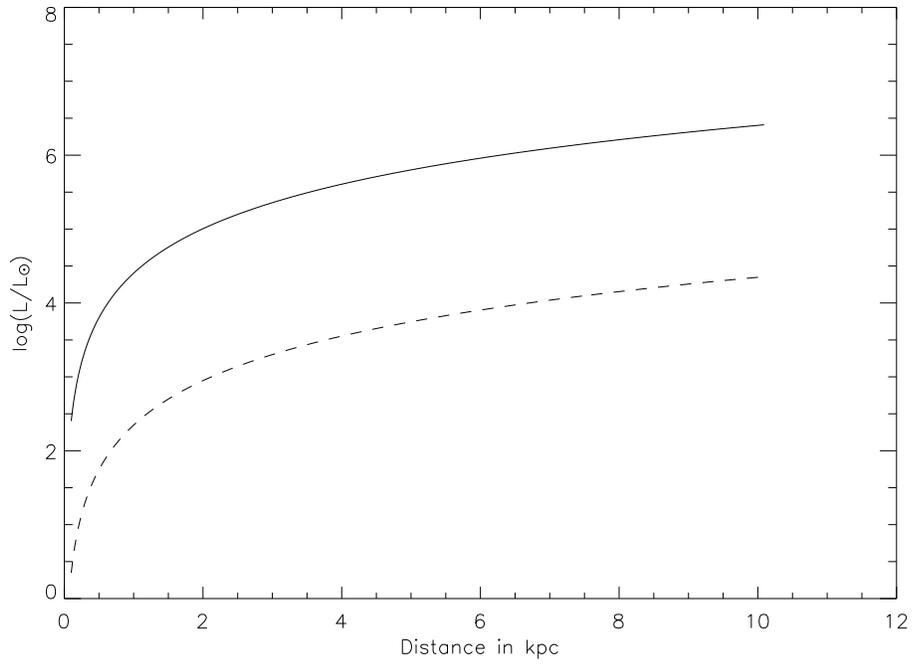}
\caption{Relation between distance and luminosity assuming $A_{V}=17.4$ (solid curve), and $A_{V}=11.8$ (dashed).}
\label{distancelum}
\end{figure}

\begin{figure}
\plotone{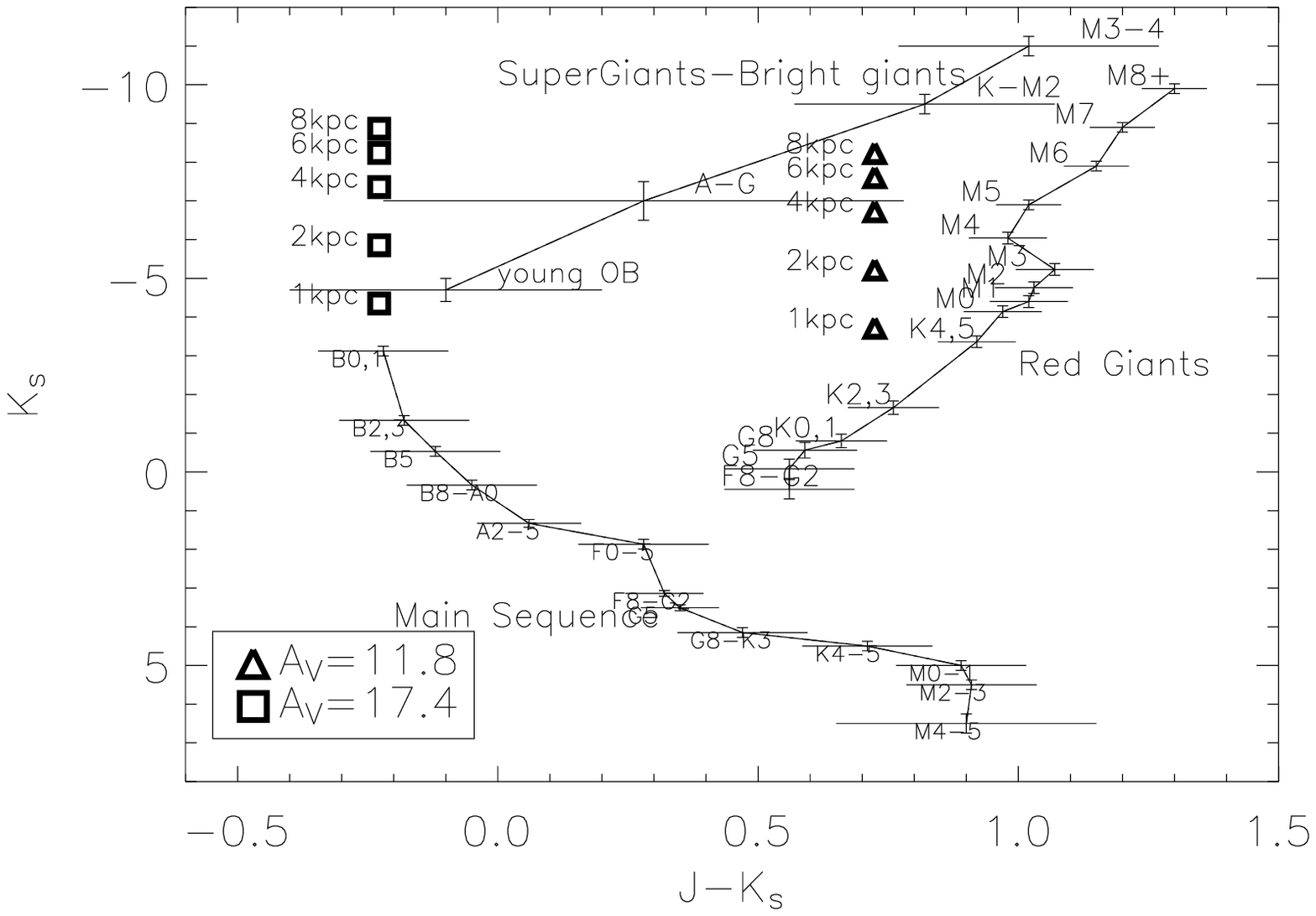}
\caption{Position on the color-color HR diagram computed from template stars of \citet{ruelas}, for distance from 1 to 8 kpc, and for two absorption values : $A_{V}=17.4$ and $A_{V}=11.8$.}
\label{diaghr}
\end{figure}

	Simple qualitative arguments show that the source is likely to be galactic, and is not an outer galaxy~:
\begin{itemize}
\item the galactic latitude is $-27\arcmin$~; usually absorption prevents observations of extragalactic objects so close to the Galactic plane (however see \citet{marti} and \citet{ribo} for two counterexamples)~; however, as INTEGRAL is aimed primarily within the galactic plane, a selection effect of unusual extragalactic objects is possible~;
\item the lines in the spectrum show no cosmological redshift~; relatively nearby objects (up to few Mpc) would show no redshift, but we can exclude the possibility of a quasar ; on the other hand, the source is point-like, with a seeing better than $1\arcsec$, in disagreement with the possibility of a nearby AGN ; 
\item in the X-rays, given the variability time scale, which leads to a limit of the distance between the X-ray source and the fluorescent emission region, the width of the Fe K$\alpha$ fluorescent line is too small to come from a Seyfert II galaxy or an Ultra Luminous Infrared Galaxy  \citep{rev1,walter}.
\end{itemize}
Estimating the distance from the $r/D$ ratio of the SED fit of Figures~\ref{specphot} and \ref{attracteur} requires making some assumptions about the source. We assume that our source is a sgB[e] star, as we will discuss in subsection \ref{massdonor}, and take the list of \citet{lamers}, giving the bolometric luminosity and the temperature for known sgB[e] stars. Using Stefan's law, we compute a star radius and use it to break the degeneracy of our fit. This gives a distance between 0.9 (Hen S35 in LMC, type B1Iab) and 6.2 kpc (R66 in LMC, type B8). As high mass stars remain usually close to star forming regions, we assume that our source is in a galactic arm. It can be then in the Sagittarius-Carina arm (0.7 kpc), the inner Scutum-Crux arm (3.2 kpc), in the Norma-Cygnus arm (4.8 kpc), and a star-forming complex at 7 kpc, where features and distances are taken from \citet{russeil}. The next intercept between the direction of the source and a galactic arm, which is the Perseus arm at 10.8 kpc, is slightly too far. Conversely, all the stars of the list give a spectrum undistinguishable from the absorbed black body fit when put at the distance and absorption of our source, but it is only a simple consistency test of the method, not a clue that our source is really a sgB[e]. However, errors are difficult to assess, because of the various parameters involved, often in a non-linear way, and also because the  characteristics of the sgB[e] stars show too much dispersion (with a possible existence of two subgroups, see \citet{lamers}) to be considered as standard candles.\\
Using the results of~\ref{discuss_specphot}, and fixing the distance and the absorption, we can have an estimate of the luminosity. This is shown on Figure~\ref{distancelum}. For $A_{V}=17.4$ (hence $T=18\,000\,\rm K$), we have $\log(L/L_{\sun})>4$ for $D>1\,\rm kpc$. The corresponding radius is $R_{\star}>10\,R_{\sun}$. These characteristics are compatible with a B0 to B5 type giant star, or B or A supergiant star. For $A_{V}=11.8$ for instance (hence $T=4\,000\,\rm K$), we have $\log(L/L_{\sun})>2$ for $D>1\,\rm kpc$. The corresponding radius is $R_{\star}>30\,R_{\sun}$. These characteristics are compatible with a G to M type of supergiant star. However we have seen that a higher absorption value is preferred by the data.\\
A consistency check can be made by putting the source on a color-color HR diagram. This is done on Figure~\ref{diaghr}, using a HR diagram computed from template stars of \citet{ruelas}, for the two absorption values considered above. The low absorption value is compatible with a distant red supergiant, or a close red giant. The high absorption case is compatible with an early-type supergiant, an OB or a B, provided that the distance is not greater than 4 kpc~; this suggests a location in the Scutum-Crux arm. The absorption cannot be substantially higher to be compatible with this figure~; this is consistent with Figure~\ref{attracteur}, showing that absorption is always below 18 for temperatures below $60\,000\,\rm K$.


\placefigure{distancelum}
\placefigure{diaghr}


	\subsection{Nature of the Source}

The fact that IGR~J16318$-$4848 was detected during an outburst of X-ray emission, and that a previous outburst was found in the ASCA archival data of 1994 \citep{murakami}, strongly points towards a binary system, including a compact object such as a neutron star or a black hole. A correlation between the X-ray and the radio emission in low/hard state black hole binaries is proposed in \citet{gallo}. According to this correlation, and given the XMM flux and absorption in the 2--10 keV band \citep{matt}, a radio flux of $43\pm 30\,\rm mJy$ is expected (where the error bar comes from the uncertainties given in \citet{gallo} for the parameters of their fit), more than one order of magnitude above the upper limit given by ATCA \citep{walter}. This is a hint that the compact object is a neutron star instead of a black hole, but given that the environment is so unique, one has to be cautious about applying this X-ray/radio relationship. Note that, similarly, the flux of CI Cam (12 mJy at 1.4 GHz according to \citet{hjellming}) is somewhat smaller that the expected flux ($52\pm 30\,\rm mJy$) given the $\sim 65\,\rm mCrab$ flux during the outbursts \citep{frontera} if the correlation of \citet{gallo} holds.\\
We detect no significant variation of the lines in wavelength, but given the short duration of the observations (10 minutes between the first and the last spectrum in each GBF and GRF bands), this is not surprising. Long duration observations would be useful in order to detect such a variation, if any, and to infer constraints on the masses of the system.\\
Using the X-ray observations by XMM and INTEGRAL, \citet{matt} and \citet{walter} infer the following unabsorbed luminosities at 1 kpc~: $1.3\times 10^{35}\rm\,erg\,s^{-1}$ between 2 and 10 keV, $3\times 10^{34}\rm \,erg\,s^{-1}$ between 20 and 50 keV. As the photon index is very close to 2, the bolometric luminosity is somewhat difficult to estimate accurately, because it strongly model dependent, in particular of the cut-off energies. Assuming the X-ray emission is dominating the SED of the compact object, we choose the sum of the two quoted values to be a good estimator of the luminosity (indeed it is a lower limit). Using the above estimation of the distance, this leads to a luminosity ranging from $1.3\times 10^{35}\rm\,erg\,s^{-1}$ (0.9 kpc) and $6.2\times 10^{36} \rm\,erg\,s^{-1}$ (6.2 kpc), at least 30 times lower than the Eddington limit of $1.8\times 10^{38}\rm\,erg\,s^{-1}$ for a neutron star of $1.4\,M_{\sun}$. That suggests that moderate accretion is taking place, but given that the environment is so unique, one has to be cautious about applying this relationship. Note that the luminosity in the X-rays during the outburst is compatible with the common values for Be/X-ray binaries, where the compact object is a neutron star \citep{negueruela}.\\
The huge difference in the absorption between the X-ray and our optical/NIR observations (the \ion{H}{1} column density is 100 times higher in the X-rays than in our observations, see \citet{walter}) indicates that the X-ray absorption is intrinsic to the compact object \citep{rev1}. It is probably due to mass loss from the companion star enshrouding the compact object, thereby absorbing the X-ray emission. Using the variability time scale of the X-ray emission, and the maximum delay observed between the Fe K$\alpha$ line and the continuum variations, \citet{walter} deduce that the distance $r$ between the zone in which fluorescent emission takes place and the X-ray source is constrained by $r<10^{13}\,{\rm cm}=140\,R_{\sun}$.
We suppose that the distance between the source of the NIR emission and the source of the X-ray emission is at least $r$ ; a hint for this hypothesis is that the absorption in the X-rays is much higher than in the NIR, although one has to be cautious, because the nature of the absorbing material might not be the same in both wavelength domains. Unless the donor is a red giant of type K or later, this excludes a symbiotic binary system~; however, we have seen that our data prefer hot stars, and therefore that a High Mass X-Ray Binary (HMXB) is preferred. Assuming that we have a neutron star of $1.4\,\rm M_{\sun}$ orbiting a B type star with $1.4\lesssim\log(M/M_{\sun})\lesssim 1.7$, and that we have $r=10^{13}\,{\rm cm}$, we obtain a lower limit for the period between 28 and 38 days. With $r=10^{12}\,{\rm cm}$, the minimum period is shortened to 0.9 to 1.2 days. In both cases, the period is far too long to have been detected by our observations. This point needs further observations to detect and measure an orbital period and would increase noticeably our knowledge of IGR~J16318$-$4848. It is possible that the X-ray outburst is close to the passage of the compact companion at the periastron, as it is the case in type I Be/X binaries \citep{negueruela}.

	\subsection{About the Mass Donor}
	\label{massdonor} 

The probably high absorption (we have seen that $A_{V}=17.4$ is favored by our data) points to a massive supergiant star, with a luminosity higher than $10^{4.4}\,L_{\sun}$ for a distance greater than $1\,\rm kpc$, $10^{5}\,L_{\sun}$ for a distance greater than $2\,\rm kpc$. The hypothesis of binarity leads to high mass X-ray binary (HMXB), the traditional mass donors of which being classical Be and OB supergiant stars. The latter, favored by the high luminosity, is not probable, because it generally leads to persistent X-ray emission.\\
The fact that the lines of the spectrum are nearly all emission lines, and the fact that among them we have forbidden lines are key constraints for the identification of the spectral type of the mass donor. Usual stars have absorption lines. The P-Cygni profiles seem to be not numerous enough to make a classification as a P-Cygni star as HDE 316285 \citep{hillier2}. Bright infrared [\ion{Fe}{2}] emission appears to be a common property among LBVs with prominent nebulae \citep{smith} ; however, the line at $1.643\,\micron$ is very bright in these cases, whereas we do not observe this line, and the other forbidden lines are weak.
Among the class of massive stars in transition, the spectral properties around $2\,\micron$ are in good agreement with both B[e] and Of/WN stars \citep{morris}. The other star types (O, B0) considered in \citet{morris} are disfavored because the Br$\gamma$ line is in absorption in these types, but as noted in this article, there is always at least one example from each group that can be easily associated with another one, and therefore robust conclusions are difficult to infer from the infrared spectrum alone. However, Wolf-Rayet stars can be ruled out, as these stars are usually very poor in hydrogen.\\
The NIR spectrum of IGR~J16318$-$4848 shows strong hydrogen Paschen, Brackett and Pfund emission lines. Although we have no spectroscopic data in the visible region, 
we can assume that Balmer emission lines are probably present at a strong level. The spectrum shows also emission lines of \ion{Fe}{2}, both allowed and forbidden transitions. The absorption seems to be higher than the estimated interstellar absorption along the line of sight (considering $A_{V}=11.8\pm 1.6$ for interstellar absorption, or considering the very unusual colors of the source in respect to the field stars, as seen on Figure~\ref{fig_mag})~; moreover, the presence of \ion{Na}{1} lines indicates the presence of circumstellar regions shielded from direct stellar radiation, but the absence of molecular $\rm H_{2}$ suggests a rather hot temperature. All this seems to converge towards the existence of hot circumstellar dust, although the lack of observations in the optical range makes difficult to assess an excess of IR emission. These points are the characteristics of the ''B[e] phenomenon'', according to the classification of \citet{lamers}.\\
The fact that the luminosity is high suggests a supergiant B[e] (sgB[e]). Some of the secondary criteria for a sgB[e] classification according to \citet{lamers} are fulfilled~: some lines show a P-Cygni profile indicating mass loss (criterion B1 of \citet{lamers}), the possibly high He/H ratio indicates an evolved star (criterion B3), although this ratio is not very firmly constrained by our data, and a high extinction (criterion B4). However, we did not see any hybrid spectrum (there is no broad absorption feature) nor strong interstellar bands (contradicting criteria B2 and B4), but the latter feature is only reported as ''usual''. The non detection of the source in the USNO B1.0 catalog suggests that photometric variations can be higher than 2 magnitudes in the $R$ band (contradicting criterion B5, saying that the variation is on the order of 0.1 to 0.2 magnitude)~; however, the 2MASS magnitude in the $J$ band (in 1999) is in agreement with our own measurements (2003). The criteria for the other B[e] subtypes (pre-main sequence stars, compact planetary nebulae or symbiotic binaries) are clearly not satisfied. In addition, as already stated, the spectrum is very similar to that of CI Cam, both in the X-ray domain \citep{boirin}, and in the NIR domain \citep{clark}. CI Cam has been classified as a sgB[e] in the optical domain by \citet{robinson} and \citet{hynes}, as the first sgB[e] associated with a HMXB. The lack of IRAS detection is a concern with this hypothesis, as CI Cam is a significant IRAS source. However, CI Cam is also much brighter than the IGR~J16318$-$4848 in the NIR and visible wavelength. By comparing the fluxes in the K band of both sources, we expect a flux at $12\,\micron$ about 1 Jy or less for IGR~J16318$-$4848, around the upper limit we estimated for IRAS. Therefore, the lack of detection is not really conclusive. 
A classification as sgB[e] appears to be a reasonable choice, even if a classification as $\rm unclB[e]/sgB[e]$ may be more secure because of the problematic fulfillment of some secondary criteria.\\ 
Therefore, we propose sgB[e] as a tentative identification for the massive companion of IGR~J16318$-$4848, making of this source the second HMXB with a sgB[e] after CI Cam. There is evidence that the circumstellar material of a sgB[e] is concentrated in a disk \citep{hubert}, with a rarefied polar wind. As already proposed by \citet{hynes} in the case of CI Cam, it is possible that the X-ray outburst is caused by the passage of the compact object in the disk, perhaps near the periastron. In this case, the burst should be periodic, as the ASCA observations in 1994 suggest \citep{murakami}.\\

\section{CONCLUSION}
\label{secconc}

The source IGR~J16318$-$4848 was the first source discovered by INTEGRAL. In the course of a ToO program using the NTT telescope, we performed  photometric and spectroscopic observations less than one month after its discovery in the optical and NIR domains. We list here our main results~: 
\begin{itemize}
\item we discovered the optical counterpart and confirmed an already proposed NIR candidate~;
\item we performed an independent astrometry for this candidate~;
\item we obtained photometric measurements for the $R$, $I$ and $J$ bands, and flux upper limits for $B$ and $V$, flux lower limits for $H$ and $K_{s}$~;
\item with the continua of our GBF and GRF spectra, these photometric measurements and with X-ray, radio and archival data, we constructed a SED covering 10 decades in wavelength~;
\item on the spectrum we identified 72 emission lines, including forbidden lines, 80~\% of these lines have been detected in CI Cam, suggesting a similar nature~;
\item the data favor an absorption of $A_{V}=17.4$, greater than the interstellar absorption, and a temperature above $10\,000\,\rm K$~;
\item the data favor the existence of a dense circumstellar material, with stellar wind~;
\item the distance is between $\sim$0.9 and 6.2 kpc~;
\item we propose as the most likely hypothesis that the source is a HMXB with a sgB[e] star as the mass donor~; it would be the second case after CI Cam~;
\end{itemize}
Complementary observations are needed in order to confirm our results, among them we propose~:
\begin{itemize}
\item high resolution NIR spectroscopy, if possible extended to optical, in order to~:
\begin{itemize}
\item extend the SED in the optical and in the mid-infrared to directly see a NIR excess~;
\item check if the similarity with CI Cam as observed in the NIR is still valid in the optical~; 
\item improve our results concerning P-Cygni profiles, and line broadening~;
\end{itemize}
\item long term follow-up spectroscopy and photometry, in order to~:
\begin{itemize}
\item seek for line variability~;
\item seek for a periodic behaviour to infer the orbital elements.
\end{itemize}
\end{itemize}
This source shows many unusual features, the first is its strong intrinsic absorption. Interestingly, among the ten sources that INTEGRAL has discovered in this region, this feature is common (at least in the X-rays), to the three sources discussed by \citet{rev2} : IGR~J16318$-$4848, IGR J16320-4751 and IGR J16358-4726, although the $N_{\rm H}$ column density is lower by an order of magnitude in the two latter systems \citep{rodriguez,patel}. However, a clear identification for the optical/NIR counterpart has been done only for IGR~J16318$-$4848. Moreover, the type of the mass donor, as inferred from our study, has been considered up to now as very rare. There is therefore the possibility that INTEGRAL, with the discovery of IGR~J16318$-$4848, has unveiled a new class of binaries that will deserve much attention in the future.

\acknowledgments

This work uses observations made at the ESO NTT upon detection and localization made by INTEGRAL and XMM. We are very grateful to the ESO staff for their availability and skills
for performing override programs, and special thanks go to Malvina Billeres
for having performed these service ToO observations.
We use the 2MASS, DENIS, USNO, GSC and DSS catalogs for astrometry and photometry. 
We acknowledge P. Ferrando, P. Goldoni, F. Lebrun, M. Rib\'{o} and J. Rodriguez for useful discussions and careful reading of the manuscript. SC thanks R. Hynes, I. Negueruela and M. Rib\'{o} for useful discussions on the nature of the source, and R. Walter for useful discussions. We thank the referee for a careful reading of the manuscript and useful suggestions. PF also acknowledge the CNRS/F\'{e}d\'{e}ration de Recherche APC for funding.

\end{document}